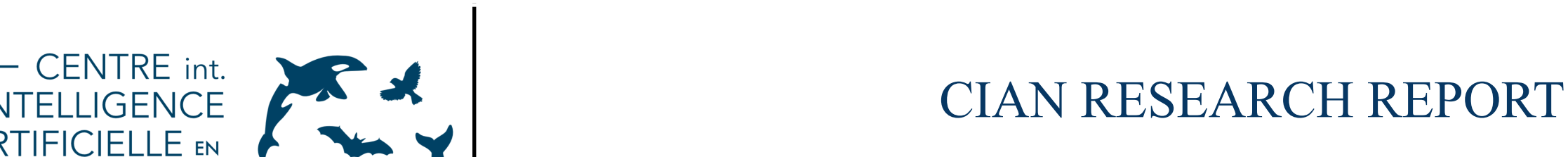
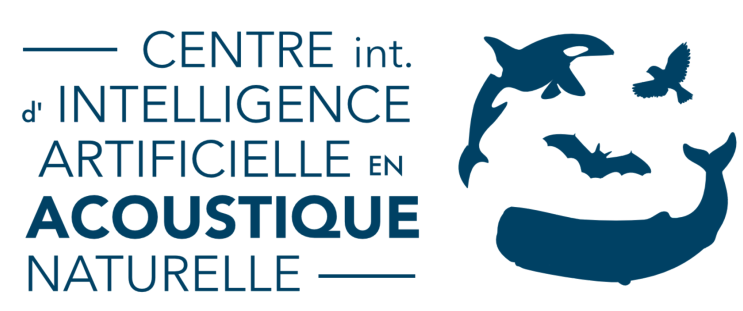

# Audiovisual Diarization of Overlapping Click Trains in Sperm Whale (*P. macrocephalus*) Vocal Sparring using a Three-Hydrophone Array

Lara Berkenbaum[1,2,6], Hervé Glotin[1,2,6], François Sarano[2,6], Walter M.X. Zimmer [3,6], Véronique Sarano[2,6], Olivier Adam[4,5,6], Pascale Giraudet[1,2,6]

1. Univ Toulon, Aix Marseille Univ, CNRS, LIS, DYNI NATAL, Toulon, France
2. Longitude 181, Valence 26000, France
3. Independent researcher, Lerici (SP) 19032, Italy
4. Jean Le Rond d'Alembert Inst., Sorbonne University, CNRS, 5005 Paris, France
5. Neurosciences Paris-Saclay Inst., Paris-Saclay Univ., CNRS, 91400 Saclay, France
6. Univ Toulon, CIAN, Toulon, France

*lara.berkenbaum@lis-lab.fr*, *glotin@univ-tln.fr*, *saranofrancois@gmail.com*, wzimmer@wmxz.eu, *veronique.sarano@longitude181.org*, *olivier.adam@sorbonne-universite.fr*, *giraudet@univ-tln.fr*

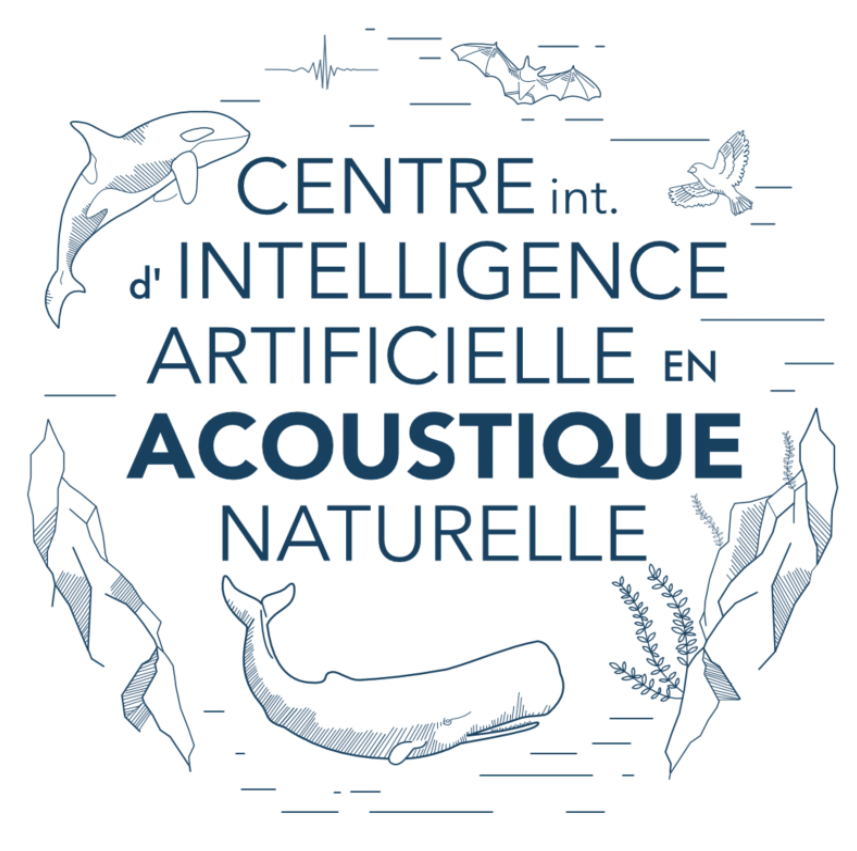

Int. Center of Artificial Intelligence in Natural Acoustics

Research Fed. RNSR 202424556S
av. de l'Université
CS 60584, 83041,Toulon, FR

https://cian.lis-lab.fr

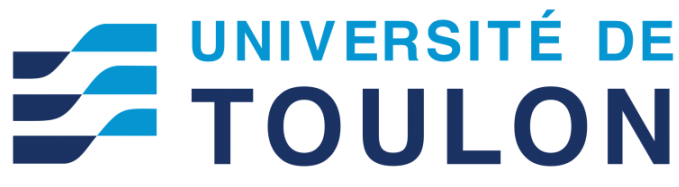

**Audiovisual diarization of overlapping click trains in sperm whale (*Physeter macrocephalus*) vocal sparring using a three-hydrophone array**

Lara Berkenbaum,[1,2,6,a] Hervé Glotin,[1,2,6,b] François Sarano,[2,6] Walter M.X. Zimmer,[3,6] Véronique Sarano,[2,6] Olivier Adam,[4,5,6] and Pascale Giraudet[1,2,6]

[1] *Univ Toulon, Aix Marseille Univ, CNRS, LIS, NATAL, Toulon, France*
[2] *Longitude 181, Valence 26000, France*
[3] *Independent researcher, Lerici (SP) 19032, Italy*
[4] *Jean Le Rond d'Alembert Institute, Sorbonne University, CNRS, 5005 Paris, France*
[5] *Neurosciences Paris-Saclay Institute, Paris-Saclay University, CNRS, 91400 Saclay, France*
[6] *Univ Toulon, CIAN, Toulon, France*

The individual attribution of sperm whale (*Physeter macrocephalus*) vocalizations during surface interactions constitutes a methodological challenge due to acoustic overlaps, multipath propagation, body shadowing, and indistinguishable inter-pulse intervals among similar-sized individuals. A multimodal audio-visual workflow is presented to deinterleave and attribute uncharacterized click trains produced by immature males during "vocal sparring" using a portable three-hydrophone array coupled with synchronized video. The approach combines spectro-temporal tracking, based on inter-click interval dynamics and the Constant-Q Transform, with spatial time-difference-of-arrival modeling projected onto the image plane for optical validation. Analysis of 2,655 manually validated clicks shows that purely acoustic clustering diarization errors remain low, peaking at 25.79% only during extreme temporal superpositions. Integrating the optical modality resolves residual spatial indeterminacies; although near-planar array geometry induces vertical ambiguities, the system achieves up to 100% horizontal visual concordance for primary emitters. Crucially, this framework successfully reconstructed and assigned 11 distinct, intertwined click trains totaling 882 clicks to specific focal individuals despite near-field tactile constraints. Because ethological descriptions remain incomplete without identifying the emitter, explicitly correlating these emissions with physical kinematics provides the fine-scale resolution required to define this socio-acoustic behavior.

[a] Email: lara.berkenbaum@lis-lab.fr
[b] Email: glotin@univ-tln.fr

# I. INTRODUCTION

While sperm whale (*Physeter macrocephalus*) acoustic communication primarily features foraging echolocation clicks (Gubnitsky et al., 2025; Madsen et al., 2002; Watwood et al., 2006) and social codas (Gero et al., 2016; Rendell and Whitehead, 2003), continuous click trains produced during surface socialization evade this dichotomy. Generated by the well-described "bent horn" mechanism (Møhl et al., 2003; Zimmer et al., 2005), these continuous emissions lack familiar rhythmic patterns. When produced during close-range, multi-emitter interactions, they create a cocktail-party effect. Consequently, assigning these signals to specific individuals, the challenge of acoustic diarization, remains a major methodological bottleneck, leaving these potentially critical socio-acoustic behaviors uncharacterized.

Isolating these emissions is essential for studying the vocal ontogeny of juvenile male sperm whales, an unmapped demographic segment within matrilineal units. Following Berkenbaum (2021), "juvenile" denotes individuals aged 0-9 years and "immature" denotes ages 4-9 years; we apply these definitions throughout, irrespective of the size- or context-based terminologies of the cited studies. Unlike adult males, whose acoustics and kinematics are well documented (Laplanche et al., 2005; Madsen et al., 2002), pre-dispersal immature males remain morphologically indistinguishable from adult females (Gero et al., 2014). Mapping their socio-cultural integration (Cantor et al., 2019; Hersh et al., 2022) requires tracking close-range interactions, which fundamentally challenges classical acoustic source separation methods and precludes reliable individual acoustic attribution.

Standard acoustic separation and localization methods fail during near-surface, multi-emitter socialization. While Time Difference of Arrival (TDOA) effectively localizes deep-diving cetaceans (Ferrari et al., 2019, 2020; Giraudet and Glotin, 2006; Glotin et al., 2008; Zimmer, 2011; Zimmer and Troiano, 2024), its spatial resolution degrades when the array aperture is small relative to the source distance. Consequently, near-field tactile interactions introduce severe geometric ambiguities for compact three-hydrophone arrays. Depth-based multipath separation (Giraudet and Glotin, 2006; Gubnitsky et al., 2025; Nosal and Neil Frazer, 2006; Thode, 2004, 2005) also fails at the surface, where the Lloyd's mirror effect (LME) merges direct and reflected paths, creating destructive frequency notches (Pereira et al., 2016, 2020). Furthermore, algorithms relying on inter-click interval (ICI) regularity (Baggenstoss, 2011; Gubnitsky et al., 2025) lose tracking precision during rapid overlaps, and waveform feature extraction suffers from dynamic on/off-axis variability (Gubnitsky et al., 2025; Møhl et al., 2003; Teloni et al., 2007).

Crucially, size-correlated inter-pulse interval (IPI) discrimination (Barile et al., 2024; Ferrari et al., 2024; Gordon, 1991) inherently fails to separate same-sized individuals. Furthermore, standard IPI frameworks typically discard near-surface signals as unreliable, and systematically assign same-aged juveniles or adult females to a single acoustic size class (Barile et al., 2024; Caruso et al., 2015). Because their near-identical IPIs render these individuals acoustically indistinguishable, their overlapping emissions are automatically conflated into a single erratic click train. Consequently, the optical modality is not merely supplementary; it is strictly necessary and serves two distinct functions. First, visual photo-identification establishes individual identity and sex, resolving the demographic ambiguity that IPI alone cannot resolve. Second, because acoustic emissions lack visible external signatures, the optical stream assigns deinterleaved clicks to specific emitters strictly through spatial concordance between the projected acoustic localization and the tracked animal's physical position. Without this dual visual grounding, acoustic data alone fail to reveal the presence of multiple emitters and the consequent need for deinterleaving.

Existing methodological innovations do not resolve these near-field challenges. Macro-scale acoustic systems (Diamant et al., 2026; Gruden et al., 2025) lack the sub-meter precision and optical ground truth required to disentangle surface interactions. Biologging (Beguš et al., 2025; Sharma et al., 2024) effectively isolates a tagged subject's emissions but is invasive, short-term, and unsuitable for identifying multiple interacting untagged individuals. While coupled camera-hydrophone arrays successfully identify vocalizing dolphins (Lopez-Marulanda et al., 2017) and other species (Best et al., 2025), applying them to sperm whales requires overcoming the combined effects of biosonar forward-directivity and surface multipath interference.

To overcome these limitations, we propose a multimodal approach combining acoustic spectro-temporal tracking, spatial localization, and optical validation to identify emitters and capture behavioral context. The spectro-temporal stage relies on the Constant-Q Transform (CQT; Brown, 1991), whose logarithmic frequency scaling is well suited to resolving the complex spectral structure of these broadband signals. Although this time-frequency representation is still emerging in bioacoustics (Himawan et al., 2018; Li et al., 2025), to our knowledge, it has not been previously applied to sperm whale clicks. Using "vocal sparring" (Berkenbaum et al., 2025a, 2025b) as a proof-of-concept, a near-field interaction involving physical contact and indistinguishable IPIs, this work aims to: (i) demonstrate a non-invasive workflow for acoustic deinterleaving and video-aided individual assignment using a mobile three-hydrophone array, designed to generalize to other same-sized individuals and socio-acoustic behaviors; and (ii) evaluate system performance using a clustering-specific diarization error rate ($DER_{clust}$) for acoustic processing, and visual concordance rates ($VCR$) to compare acoustic localizations against tracked subjects, thereby explicitly quantifying the optical modality's contribution to resolving complex socio-acoustic scenes. Beyond validating the workflow, assigning these overlapping click trains to identified individuals is the necessary condition for confirming that the emissions co-occurring with vocal sparring are indeed produced by the interacting whales, an association that cannot be established by acoustic co-occurrence alone.

## II. MATERIALS AND METHODS

### A. Study context and subjects

Data originate from long-term opportunistic underwater observations off the coast of Mauritius (Indian Ocean), conducted as part of the "La Voix des Cachalots" program (2015-2024; CIAN Institute; Longitude 181; Un Océan De Vie).

This study focuses on a specific socio-acoustic behavior known as "vocal sparring" (Berkenbaum et al., 2025a, 2025b), defined as non-aggressive, voluntary dyadic interactions combining close physical contact with sustained overlapping click trains. These interactions feature a stereotyped suite of physical kinematics, including varied tactile contacts (mouth-to-mouth, mouth-to-body, head-to-head, body-to-body), rolls, and approach-retreat sequences, along with mouth opening and closing. These kinematic sequences co-occur with continuous clicking, silences, and occasionally followed by fast trains ("buzzes"). Within the monitored social group, vocal sparring occurs exclusively among same-age juvenile male dyads. It meets the primary ethological criteria for social play (Bekoff, 1972; Burghardt, 2014) and likely contributes to pre-dispersal vocal ontogeny and social bond formation. Here, we analyze interactions between two five-year-old immature males, born in February and April 2018, identified as Ali and Daren using the local population photo-identification catalog (Sarano et al., 2022).

While standard single-sensor recordings of these events are abundant, multimodal data capable of resolving individual emitters are scarce. To overcome this limitation, we collected data during a two-week deployment of the compact three-hydrophone OPALE array (Fig. 1) in 2023. This configuration provides the spatial resolution needed to help separate severe acoustic overlaps and establish a reliable baseline for individual acoustic assignment. From the 2023 field campaign, we selected four video sequences containing vocal sparring synchronized with eight audio recordings (each 96.54 s in duration, recorded on 4 May 2023; Sec. II.B for detailed synchronization methodology).

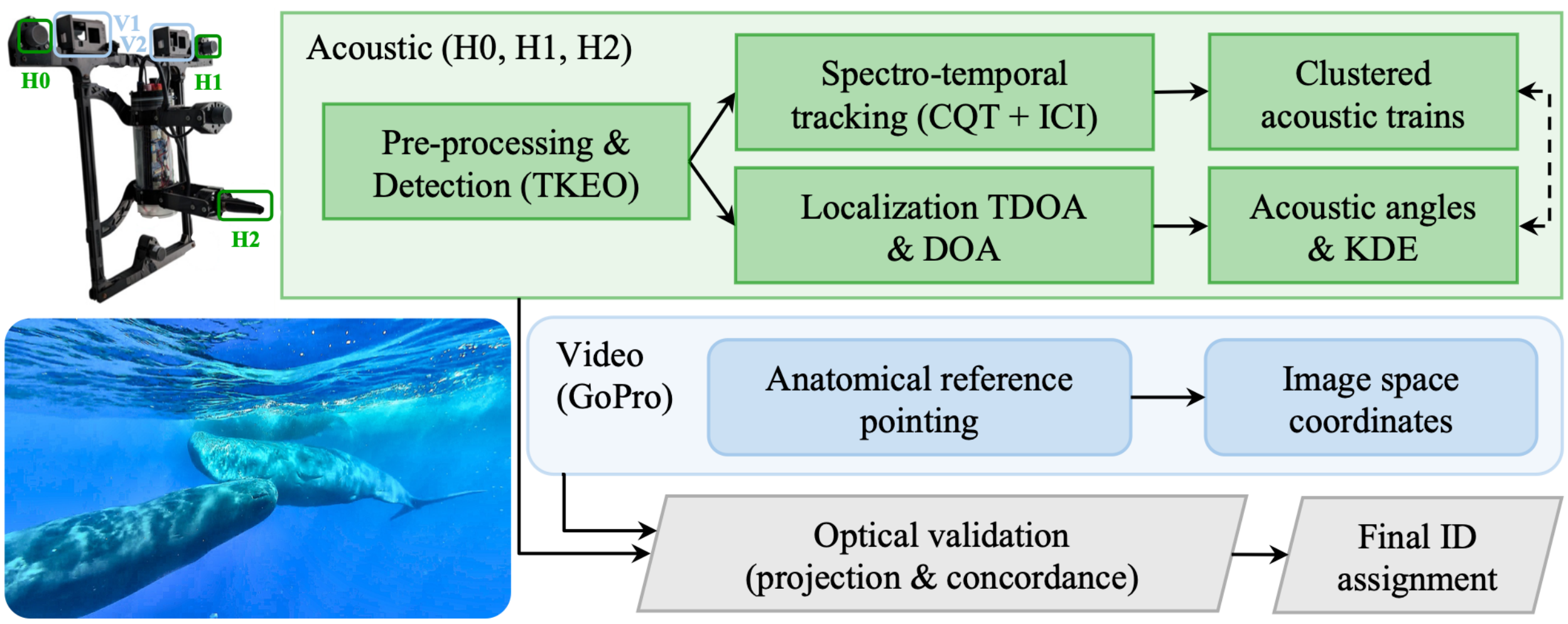

FIG. 1. (Color online). Multimodal system for acoustic diarization during close-range vocal sparring. The left panel display the compact OPALE audio-visual array and a contextual surface interaction frame involving multiple individuals. The right panel illustrates the processing architecture flowchart, detailing the parallel acoustic tracking and spatial localization, independent video data processing, and their integration to validate projected acoustic angles and assign speaker identity.

### B. Audio-visual acquisition system: OPALE array

We used the manually deployed compact OPALE array, a passive acoustic monitoring (PAM) system developed by the Intelligent Acoustics team (CIAN, DYNI; Ferrari et al., 2020; Glotin et al., 2024; Intelligent Acoustics, 2026), to ensure non-invasive data acquisition. A scientific swimmer maintained the system at a maximum depth of 5 m, ensuring a nominal individual-to-array distance of 10-30 m. All field surveys were conducted in the morning in accordance with the Republic of Mauritius Maritime Zones (Conduct of Marine Scientific Research) Regulations 2017 (Government Notice No. 57 of 2017) and followed the official charter for responsible approach protocols of marine mammals, ensuring no avoidance behavior was induced.

A rigid 3D-printed frame enables simultaneous audiovisual recording and guarantees consistent geometry (Fig. 1). Due to hardware constraints during the 2023 deployment, the system was limited to a functional sub-array of three heterogeneous sensors. Channels 0 and 1 ($H_0$ and $H_1$; SQ26-H1B) define the main horizontal axis, while channel 4 ($H_2$; C75) is positioned lower and forward to form the inclined detection plane necessary for 3D spatial resolution. Detailed sensor specifications and the exact coordinate matrix defining the array geometry are provided in Supplementary Material (S1). The on-board acquisition system digitized acoustic signals at a sampling rate of 256 kHz (24-bit resolution) to capture the full spectral content of the sperm whale clicks.

The array integrates two GoPro Hero 10 cameras (4K, 60 fps) mounted on the horizontal plane. To preserve geometric correspondence between optics and acoustics, we disabled optical distortion (fisheye) and automatic stabilization, yielding a field of view (FOV) of 118° horizontally and 69° vertically. This provides the angular coverage necessary to capture close-range surface interactions within the array's detection range. The internal QHB clock ensured hardware synchronization of the acoustic channels, and we retrospectively aligned the audio and video streams via cross-correlation on a shared transient marker (an out-of-water clap) and refined by visually matching the recorded clicks across streams.

After a preliminary visual inspection of the dataset to isolate occurrences of "vocal sparring", a subsample comprising three recordings (files 081356, 085315 and 085452) was selected for in-depth analysis of acoustic diarization and spatial localization. These sequences capture the exact moments the focal individuals were closest to the array, engaging in the targeted socio-acoustic behavior, and clearly identifiable in the video stream. We subdivided these sequences into 12 distinct interaction scenes. To objectively quantify apparent acoustic complexity prior to deinterleaving, we computed the coefficient of variation (CV) of the raw ICIs for all 12 scenes (Table S2), distinguishing simple click trains from those containing high degree of overlap. Three scenes were subsequently excluded: two because the primary focal subjects temporally exited the camera field of view during highly complex acoustic sequences (elevated CV; Scenes 10 and 11), and one owing to incomplete click detection precluding reliable clustering (Scene 12). The remaining nine focal scenes were retained for the core analysis.

### C. Acoustic pre-processing and click detection

To accurately compute TDOA across a heterogeneous array, we first relatively calibrated the channels to physical pressure (µPa) (Gillespie et al., 2020; Zimmer, 2011). Detailed hardware sensitivities, empirical corrections, and polarity inversion parameters are provided in the Supplementary Material S2. Consistent with the literature (Urick, 1983; Wenz, 1962), raw signal analysis revealed a predominance of low-frequency oceanic noise masking the acoustic transients, particularly on the broadband C75 sensor. To mitigate this masking, we applied a fourth-order Butterworth high-pass filter (5-kHz cut-off frequency), yielding a median signal-to-noise ratio (SNR) of 41.8 dB (SQ26) and 38.85 dB (C75). We designated Channel 0 (SQ26) as the reference hydrophone to maintain horizontal symmetry and optimize cross-correlation accuracy (Zimmer, 2011).

Following filtration, we extracted impulse energy using the discrete nonlinear Teager-Kaiser Energy Operator (TKEO; Giraudet and Glotin, 2006; Glotin et al., 2008; Kaiser, 1990; Kandia and Stylianou, 2006; Zimmer, 2011), Eq. (1):

$$\Psi(x[n]) = x^2[n] - x[n+1]x[n-1] \tag{1}$$

To mitigate passive acoustic background noise variability, the TKEO trace was adaptively normalized via a sliding-window local median and smoothed using a recursive leaky integrator (Gillespie et al 2013, Mellinger 2007, Zimmer 2011; full parameters in Supplementary Material S2). Processing dense social interactions presents a fundamental trade-off: continuous track clustering requires high sensitivity, whereas robust spatial localization (TDOA) demands steep-front signals to prevent spatial aberrations (i.e., 'ghost whales') (Zimmer, 2011). To address this, we implemented a two-stage asymmetric detection strategy. First, to maximize track continuity for spectro-temporal clustering, we applied permissive TKEO detection thresholds. Conversely, to guarantee the mathematical precision of TDOA estimates for spatial inference, we enforced strict thresholds ensuring a near-zero constant false alarm rate (CFAR). All detection thresholds were optimized via Receiver Operating Characteristic (ROC) curve analysis (Fawcett, 2006), with exact values detailed in the Supplementary Material S3.

We manually validated all automated detections, strictly defining impulse onset and offset times to discard residual noise, a fundamental step for accurate indexing and establishing the ground truth for our $DER_{clust}$. Next, we applied an adaptive envelope tracking filter (parameters in Supplementary S2) to account for focal individual movements and exclude distant biological choruses.

The CV of the ICI, used to classify the rhythmic stability of these environments, is defined in Eq. (2):

$$CV = \frac{\sigma}{\mu} \tag{2}$$

We empirically set a threshold of $CV \geq 0.5$ to formally distinguish low-variance single-source emissions from highly overlapping, multi-source socio-acoustic environments, thereby dictating the required algorithmic complexity for subsequent source separation (Table S2).

Following detection, the clicks feed two parallel processing chains with distinct objectives and, consequently, distinct signal-conditioning and sensor requirements (Fig. 1). The clustering chain (Sec. II.D) segregates individual click trains from their spectral signatures; it operates on a single reference hydrophone (SQ26) and applies a derivative-based pre-whitening to the raw waveform, prioritizing spectral detail. The localization chain (Sec. II.E) estimates the spatial origin of each click; it operates on all three hydrophones and uses the 5-kHz high-pass-filtered signal, prioritizing sharp wavefronts for accurate TDOA estimation. These divergent requirements explain why the two chains apply different pre-processing: the clustering chain rebalances the spectral envelope to expose fine structure, whereas the localization chain preserves clean impulse onsets across the array.

### D. Spectro-temporal tracking and acoustic clustering

We tailored a dual-stage signal conditioning framework specifically for the clustering pipeline, utilizing the SQ26 reference hydrophone (channel 0) to segregate individual sperm whale click trains. Ambient marine noise and hydrodynamic flow typically dominate the lower spectrum of these surface deployments, compounding the intrinsic $1/f$ spectral tilt of sperm whale signals.

To counteract this instrument-specific noise and spectral decay, we applied a first-order discrete-time derivative operator, Eq. (3), exclusively to the digitized acoustic pressure waveform prior to spectral extraction:

$$x[n] = s[n] - s[n-1] \tag{3}$$

This discrete differentiation acts as a pre-emphasis, pre-whitening filter with a +6 dB/octave gain. It flattens the $1/f$ ocean noise profile, redistributing energy so that the high-frequency transient components necessary for distinguishing concurrent click trains within a scene are relatively enhanced with respect to the low-frequency energy that otherwise dominates the spectrum. Importantly, this operation rebalances the spectral envelope rather than removing any frequency band: the low-frequency content is retained, not suppressed. We carefully calibrated this conditioning to the SQ26 hardware profile, as applying an identical gain to flat-response, broadband hydrophones (e.g., the C75 sensor) could artificially over-amplify extreme high-frequency noise and degrade spectral resolution.

While the Short-Time Fourier Transform (STFT) remains the conventional standard in marine bioacoustics, the CQT provides an adaptive dynamic resolution via its logarithmic frequency binning. Crucially, it preserves the high spectral resolution in lower frequency bands to accurately map environmental interference patterns (e.g., Lloyd's mirror notches). To circumvent the CQT's inherent trade-off between low-frequency spectral resolution and absolute temporal precision (Li et al., 2025), our framework explicitly decouples temporal analysis from acoustic clustering; absolute temporal precision is derived from the initial TKEO detections, whereas the CQT acts solely as a dedicated spectro-temporal tracking engine to group individual click trains.

Because surface-reflection effects distribute geometry-dependent high-frequency content across the spectrum, content made accessible by the pre-whitening, we computed the CQT across a wide band from 1.2 kHz to 80 kHz. The 1.2-kHz lower limit excludes dominant oceanic noise while retaining essential low-frequency spectral structure. The 80-kHz upper limit, well within the Nyquist limit, captures this geometry-dependent content, which helps separate concurrent click trains within a scene. To correct temporal imprecision from the initial detection, each click was finely realigned to its absolute energy peak before extracting and standardizing a symmetric time-frequency patch. This standardization mitigates absolute gain effects linked to propagation distance, yielding a spectral signature per click (Stowell, 2022). Detailed extraction parameters are provided in the Supplementary Material S4.

To deinterleave continuous signal overlaps, an adaptive time-frequency tracking algorithm reconstructed click trains by combining spectral validation and predictive rhythmic constraints (Baggenstoss, 2011; Nosal and Neil Frazer, 2006). Candidate click assignment relied on a weighted similarity score and a dual-memory system that compared each candidate to both the dynamically updated average profile of the emitter and the last validated click in the track. This dual memory accommodated spectral variability induced by changing orientation (Møhl et al., 2003; Teloni et al., 2007; Zimmer et al., 2005) and surface reflections, while the CQT's logarithmic scale preserved cosine distance during linear frequency shifts (Stowell, 2022).

Concurrently, a predictive rhythmic constraint exploited the short-term inertia of the ICI (Baggenstoss, 2011). Deviations from a dynamically updated temporal acceptance window incurred a penalty to the overall assignment score (algorithmic thresholds, tolerances, and scoring metrics are detailed in the Supplementary S4). Finally, an *a posteriori* fusion phase reconnected track fragments artificially interrupted by temporal impulse superposition, verifying global rhythmic consistency and spectral similarity at junction points. The full implementation of the tracking and fusion algorithm is provided in the accompanying repository (see Data Availability).

### E. Acoustic spatial localization and directional estimation

In parallel with spectro-temporal tracking (Fig. 1), the TDOA (denoted $\tau$) between each hydrophone pair was estimated by cross-correlation in the frequency domain (Knapp and Carter, 1976; Zimmer, 2011), using the sensor geometry defined by the coordinate matrix $\mathbf{H}$ (Eq. S1). To overcome quantization limits dictated by the sampling frequency, we applied quadratic interpolation to the cross-correlation peak, yielding sub-sample temporal precision crucial for a compact array. We converted the temporal delay vector $\tau$ into an apparent distance vector $b = \tau c$, where $c$ = 1541 m/s is the local sound speed, estimated from in-situ temperature, salinity, and depth conditions at the study site.

To ensure unambiguous localization in this close-range social context, we applied stringent geometric validity thresholds ($\leq 10cm$) to both the temporal loop closure error and the geometric spatial residual. This dual filtering validates temporal consistency and mitigates surface multipath effects or sensor phase heterogeneity (Nosal and Neil Frazer, 2006; Wahlberg et al., 2001; Zimmer, 2011). Although this strict thresholding rejects heavily distorted clicks, the prior spectro-temporal clustering enables reliable spatial assignment of an entire sequence based strictly on its most robust directional impulses.

Direction of Arrival (DOA) estimation relied on the plane wave assumption, validated by a source distance (10-30 m) significantly exceeding the array aperture (0.5 m). Because the degraded three-hydrophone sub-array is constrained to an inclined two-dimensional plane, it lacks volumetric rank. Consequently, we applied a constrained two-dimensional directional approach (Zimmer, 2011). We projected the geometry onto the sensor plane and estimated the tangential directional vector $\mathbf{g}_{2D}$ using the Moore-Penrose pseudo-inverse ($\mathbf{D}^{+}$), providing the optimal least-squares solution (Spiesberger and Wahlberg, 2002), Eq. (4):

$$\mathbf{g}_{2D} = \mathbf{D}^{+}\mathbf{b} \tag{4}$$

We subsequently reconstructed the missing normal component ($g_n$) by exploiting the unit norm property of the direction vector (Zimmer, 2011), Eq. (5):

$$g_n = \sqrt{1 - \|\mathbf{g}_{2D}\|^2} \tag{5}$$

When the estimated tangential vector marginally exceeded the unit-norm constraint, the normal component was recovered by taking the absolute value of the radicand in Eq. (5), which folds the estimated direction back onto the cone boundary rather than discarding the click. Since Eq. (5) as written assumes a non-negative radicand, this absolute value is applied to accommodate these marginal cases.

Equation (5) inherently yields two solutions ($\pm g_n$), creating a front/back spatial ambiguity. We resolved this indeterminacy by enforcing the positive root ($g_n > 0$, frontal half-space). This operational postulate is supported by the physical acoustic shadowing of the array frame and the directivity loss of the SQ26 sensors above 10 kHz. Definitive validation of this frontal assumption occurs retrospectively: only a spatial assignment stably intersecting a visual target during optical data fusion definitively excludes symmetrical 'ghost' locations.

Finally, to spatially validate the clustered trains, we corrected these calculated acoustic angles (azimuth and elevation) using a rotation matrix aligning with the camera's frontal reference frame (Zimmer, 2011). We applied a probabilistic Kernel Density Estimation (KDE) (Worton, 1989) within this aligned space to identify the dominant emission axis (mode) and spatial concentration zones for each train (Møhl et al., 2003). Isolating these modes geometrically verifies that distinct acoustic clusters possess resolvable spatial distributions, thereby validating the upstream spectro-temporal tracking.

### F. Optical validation and performance metrics

To fuse acoustic and optical data, we expressed the reconstructed 3D direction vector in the array's inclined reference frame. To superimpose these detections onto the optical field, we aligned the spatial geometry with the camera's orthonormal reference frame using a 3D rotation matrix $\mathbf{R}_y(-\beta_x)$ about the lateral y-axis, where $\beta_x$ is the oblique sensor's physical inclination angle (Zimmer, 2011). We then extracted the azimuth ($\theta$) and elevation ($\phi$) relative to the optical axis using spherical trigonometry (Thode, 2004; Zimmer, 2011).

Projecting these aligned coordinates onto the synchronized video's 2D image plane, bounded by its 118° x 69° FOV (Sec. II.B), establishes an independent optical ground truth. This resolves the front/back spatial ambiguities inherent in the compact array. Given the prior acoustic clustering, confirming that a subset of precise localizations intersects a target animal definitively assigns the entire click train to that emitter (Fig. 1). We finalized these assignments through frame-by-frame visual inspection, ultimately linking the deinterleaved acoustic sequences to the observed physical behaviors. Within each multi-emitter scene, the emitter contributing the greater number of clicks is designated the main speaker and the other the secondary speaker; single-emitter scenes contain a single track. The projection procedure and the resulting per-scene assignment videos are available in the repository (see Data Availability).

To quantitatively evaluate the multimodal system's performance, we defined two metrics:

$DER_{clust}$: to assess clustering accuracy independently of detection performance, we manually validated all automated detections prior to analysis, eliminating false positives and minimizing missed detections by design. Consequently, $DER_{clust}$ quantifies assignment errors, speaker confusions and omissions, normalized by the total number of manually validated clicks in the scene (including non-focal emissions), Eq. (6):

$$DER_{\text{clust}} = \frac{N_{\text{confusion}} + N_{\text{omission}}}{N_{\text{total}}} \times 100 \tag{6}$$

where $N_{\text{confusion}}$ denotes clicks assigned to the wrong emitter, $N_{\text{omission}}$ denotes validated clicks left unassigned, and $N_{\text{total}}$ represents the total number of manually validated clicks in the scene. Each acoustic cluster was associated with its majority validated emitter (a many-to-one assignment), meaning that any click, focal or non-focal, departing from its cluster's dominant identity was counted as a confusion. Under this convention, $DER_{clust}$ quantifies within-cluster purity and unassigned validated clicks; it does not penalize over-segmentation, whereby a single emitter may be distributed across several coherent clusters that are each correctly attributed under the majority rule. Unlike the standard Diarization Error Rate (DER), which incorporates missed speech and false alarm (Anguera et al., 2012), our metric excludes detection errors by design. Therefore, reported $DER_{clust}$ values represent a clustering-specific performance bound and are not directly comparable to standard speech DER.

$VCR$: To quantify spatial superposition accuracy, we manually tracked the distal end of the "junk", the primary radiating surface of the sperm whale's directional biosonar (Møhl et al., 2003; Zimmer et al., 2005), as the anatomical reference point using 2D image coordinates (Fig. 1, blue). The $VCR$ represents the percentage of projected acoustic localizations falling within a 10° spatial tolerance threshold around this anatomical reference, evaluated separately along the horizontal ($VCR_H$) and vertical ($VCR_V$) axes of the image plane angular coordinate system. The 10° tolerance was selected as the minimum value preserving concordance across all interaction contexts while remaining spatially discriminant (Sec. III. C).

To further characterize spatial projection errors and identify distinct operational regimes, we applied an unsupervised Gaussian Mixture Model (GMM; Reynolds, 2009) to the two-dimensional relative projection errors of the spatially filtered clicks subset. We computed these errors as the angular deviation (in degrees) between each projected acoustic localization and the manually tracked anatomical reference point. This probabilistic clustering decomposes the error distribution into physically interpretable components without imposing prior assumptions on cluster geometry. We determined the optimal number of components by jointly applying the Bayesian Information Criterion (BIC; Schwarz, 1978) and a solution stability analysis across 50 random initializations, quantified by the mean Adjusted Rand Index (ARI; Hubert and Arabie, 1985). Finally, we interpreted the GMM components using acoustic and kinematic metadata to identify the physical origins of each error regime.

## III. RESULTS

### A. Dataset validation and overall acoustic complexity

Our analysis includes three audiovisual recording sessions (081356, 085315, and 085452) selected during preliminary behavioral triage (Sec. II.A), totaling 4.5 minutes of synchronized data. To establish a robust acoustic reference for performance evaluation, we manually annotated 3,161 automatic detections. This yielded a validated ground-truth dataset of 2,655 true sperm whale clicks (850, 1,295, and 510 clicks for sessions 081356, 085315, and 085452, respectively). Consistent with our two-stage detection strategy (Sec. II.C), prioritizing high sensitivity to maximize track continuity during dense overlaps yielded a 96.16% recall rate (102 false negatives). As expected, this came at the expense of precision (80.77%; 608 false positives, primarily surface noise). We discarded these false positives during manual validation to ensure a pristine dataset for subsequent spatial processing (Table S1).

Following the selection criteria established in Sec. II.B., our analysis focused on the nine retained interaction scenes, containing a total of 1,368 clicks. To objectively quantify the raw acoustic complexity of these environments without bias, we computed the ICI CV for each scene using this unfiltered pool (Table I). Although the focal subjects are present throughout the dataset, acoustic overlap density varies significantly, reflecting a broad behavioral gradient. This ranges from a simple, single-emitter sequence (Scene 7; $CV = 0.152$) to moderately complex scenes featuring a focal subject vocalizing alongside isolated background events (Scenes 1 and 2; $\mathrm{CV} \approx 0.49$). The dataset culminates in highly entangled, tactile "vocal sparring" interactions characterized by sustained, simultaneous acoustic superposition between both focal subjects (Scenes 3, 4, 5, 6, 8, and 9; CV ranging from 0.567 to 0.720).

TABLE I. Total detected true-positive clicks and inter-click interval (ICI) coefficient of variation (CV) for the nine focal interaction scenes. Scenes with a $\mathrm{CV} \geq 0.5$ (bold) were classified as multi-emitter socio-acoustic environments.

| **Session** | 081356 | 081356 | 081356 | 081356 | 085315 | 085315 | 085452 | 085452 | 085452 |
|---|---|---|---|---|---|---|---|---|---|
| **Scene** | 1 | 2 | 3 | 4 | 5 | 6 | 7 | 8 | 9 |
| **Clicks** | 104 | 70 | 192 | 289 | 123 | 188 | 82 | 157 | 163 |
| **CV** | 0.487 | 0.494 | **0.567** | **0.720** | **0.581** | **0.603** | 0.152 | **0.697** | **0.697** |

To prepare the data, we applied the adaptive envelope tracking filter (Sec. II.C). While this energy threshold was computed globally across the entire recording sessions (reducing the overall 2,655 validated clicks to 2,319), applying this filter mask specifically to the nine focal scenes excluded 126 low-energy clicks. This step ultimately yielded the final baseline pool of 1,242 clicks, comprising both focal and non-focal emissions, across 2.25 minutes of cumulative acoustic activity (Appendix Table IV). Renumbered chronologically by session, then by ascending acoustic complexity within each session, these 1,242 signals form the baseline acoustic pool for evaluating the source separation and attribution pipeline.

### B. Acoustic clustering and deinterleaving performance

The baseline acoustic algorithm, relying sequentially on CQT spectral extraction and ICI rhythmic tracking, isolated overlapping click trains across a broad complexity gradient. Table II details the track-level acoustic clustering performance for the focal interactions. The algorithm achieved flawless clustering, yielding a scene-level $DER_{clust}$ (Eq. 6) of 0.0% across simple and moderately complex scenes (Scenes 1, 2, and 7). However, performance progressively degraded in heavily entangled multi-emitter scenarios (e.g., Scenes 3, 4, 8 and 9); perfect temporal superpositions introduced baseline errors, exposing the fundamental limits of standard acoustic features in near-surface, close-range interactions. For instance, Scene 8 exhibited a scene-level $DER_{clust}$ of 5.59% and Scene 9 exhibited 6.98%, while the severe acoustic masking in Scene 3 resulted in an 8.78% error rate. The most complex interaction, Scene 4, reached a 25.79% $DER_{clust}$. Resolving these specific overlaps to achieve final separation ultimately required integrating energy differentials and track junction consistency.

TABLE II. Track-level acoustic clustering performance across the focal interaction scenes. For each track, recall is the percentage of clicks correctly assigned to the track's majority emitter under the many-to-one convention used for Diarization Error Rate clustering ($DER_{clust}$; Eq.6); it is the per-track complement of the confusion errors aggregated in $DER_{clust}$. Mean ICI denotes the mean inter-click interval. *Note.* Scenes 5 and 6 are excluded from this table as their fragmented preliminary tracks could not be definitively assigned to the focal individuals due to subsequent spatial ambiguities.

| Session | Scene | Track | Role | Total clicks | Mean ICI (s) | Track recall % |
|---|---|---|---|---|---|---|
| 081356 | 1 | 1 | Main | 76 | 0.213 | 100% |
| 081356 | 2 | 2 | Main | 48 | 0.230 | 100% |
| 081356 | 3 | 3 | Main | 102 | 0.117 | 95.10% |
| 081356 | 3 | 4 | Secondary | 95 | 0.138 | 94.74% |
| 081356 | 4 | 5 | Main | 142 | 0.124 | 80.99% |
| 081356 | 4 | 6 | Secondary | 74 | 0.133 | 62.16% |
| 085452 | 7 | 7 | Single | 82 | 0.104 | 100% |
| 085452 | 8 | 8 | Main | 82 | 0.143 | 92.68% |
| 085452 | 8 | 9 | Secondary | 53 | 0.309 | 96.08% |
| 085452 | 9 | 10 | Main | 79 | 0.132 | 89.87% |
| 085452 | 9 | 11 | Secondary | 49 | 0.183 | 100.00% |
| Total | | | | 882 | | |

To dissect the algorithm's performance and understand the errors summarized in Table II, we examined the individual behaviors of its core metrics: spectral similarity and temporal rhythm. Spectrally, the CQT maintains high intra-train consistency, effectively handling intrinsic low-frequency variations (e.g., below 2 kHz) while the dominant acoustic energy remains concentrated between 5 and 20 kHz.

To accurately observe the physical impact of the propagation environment, we examined raw CQT profiles without applying the pre-whitening discrete derivative filter (Sec. II. D; Fig. 2). This raw spectral signature exhibits high broad-scale variability between different socio-acoustic contexts. For instance, Scene 1 demonstrates complex spectral structuring with pronounced frequency notching, whereas Scene 7 exhibits a more continuous, broadband energy distribution. These spectral disparities highlight the extreme sensitivity of raw acoustic features to the highly dynamic near-surface environment, a complex interplay of array positioning, animal kinematics, and multipath propagation requiring careful interpretation.

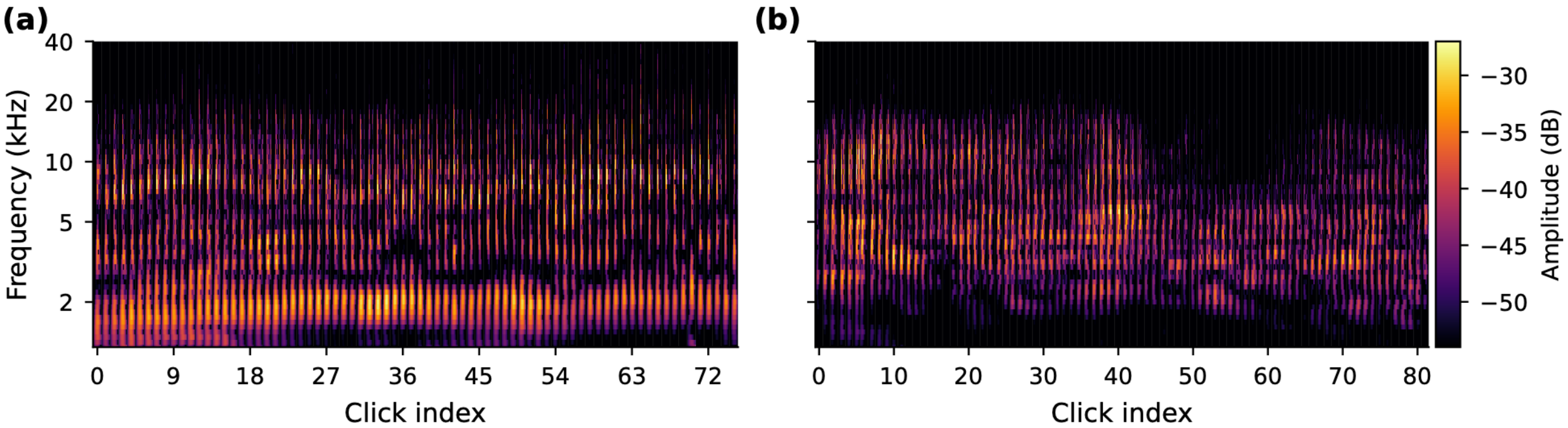

FIG. 2. (Color online). Raw Constant-Q Transform (CQT) spectrograms highlighting the spectral structure of the main speaker's clicks for (a) Scene 1, Track 1 and (b) Scene 7, Track 7. Note that the x-axes are indexed by click number rather than absolute time, allowing for a comparative spectral analysis decoupled from inter-click interval (ICI) variations. The color scale in panel (b) applies to both panels. Note the pronounced broadband continuity in Scene 7 compared to the heavily structured, multipath-induced frequency notching observed below 2 kHz in Scene 1.

Conversely, reconstructing continuous sequences within the automated workflow requires applying the pre-emphasis derivative filter to flatten the environmental spectral tilt, thereby sharpening acoustic transients and yielding robust similarity scores. Overall, the algorithm maintained strong intra-track CQT similarity scores (0.63 to 0.87) across the dataset. However, despite high efficacy in unmasked conditions (peak mean similarity of 0.87 for Track 1, Scene 1), the CQT inherently loses discriminative power during perfect temporal superpositions.

In highly entangled interactions (e.g., Scene 3, Fig. 5), overlapping impulses severely deformed the weaker track's extracted CQT patches. Consequently, while primary Track 3 maintained a robust mean similarity of 0.80, Track 4's structural deformation skewed its mean similarity down to 0.63. Furthermore, the elevated inter-track similarity (0.66) between these two specific emitters directly challenges spectral discrimination, explaining the residual clustering errors observed in this scene (8.78% $DER_{clust}$).

Alongside spectral similarity, the algorithm relies on temporal features to reconstruct coherent sequences. Across the retained dataset, mean track ICIs ranged from 0.10 to 0.31 s (Table II). While the ICI constitutes a robust primary segregation criterion when emission rhythms differ significantly (e.g., Scene 8), it loses stability during partial temporal superpositions caused by similar inter-emitter rhythms or highly variable intra-train decelerations (e.g., Scene 1). For example, Scene 4's erratic ICI severely degraded temporal tracking, directly driving the scene-level $DER_{clust}$ to a dataset peak of 25.79%, even though robust CQT similarity scores ultimately aided final track reconstruction. In Scene 9, the temporal proximity of overlapping tracks (mean ICIs of 0.132 and 0.183 s) challenged ICI-based clustering, contributing to a 6.98% scene-level error rate. However, global energy differentials (-24.0 to -35.2 dB across all tracks) resolved these ambiguities as a secondary segregation criterion. Conversely, in the highly entangled Scene 3, both temporal and energetic features lost discriminative power due to similar rhythms (ICIs of 0.117 and 0.138 s) and nearly identical received energies (-25.6 and -26.85 dB). Because energy variations are sensitive to off-axis effects and physical body masking, this combined failure left the algorithm entirely dependent on the CQT, which suffered from mathematical masking, yielding the 8.78% $DER_{clust}$.

Regarding the IPI, while estimable in single-emitter contexts (yielding approximately 2.9 ms in Scene 7, consistent with the subjects' juvenile stage (Ferrari et al., 2024), broader computational attempts on overlapping tracks yielded aberrant or bimodal distributions. Mirroring the broad-scale variability observed in the raw CQT profiles, these IPI extraction failures correlated with the animals' operating depth. Immediate surface interactions frequently produced aberrant, unreadable pulse structures, whereas sub-surface interactions predominantly yielded artifactual bimodal distributions. Compounded by the focal subjects' identical morphology (producing indistinguishable theorical IPIs), the IPI proved unreliable as a segregation metric for this dataset.

Two highly entangled multi-emitter environments, Scenes 5 and 6 (CV of 0.581 and 0.603, respectively; Table I), tested the algorithm's limits. Despite severe signal saturation, the algorithm segregated most preliminary click trains. In Scene 5 (110 total clicks), the system extracted two primary tracks accounting for 92 clicks while strong spectral coherence: the main sequence (55 clicks) yielded a median CQT similarity of 0.896, and the secondary sequence (37 clicks) reached 0.822. Similarly, in Scene 6 (185 total clicks), it extracted four distinct sub-tracks (68, 32, 27, and 21 clicks) with mean CQT similarities between 0.71 to 0.87.

These complex interactions, however, highlight the boundary conditions of purely acoustic clustering. As acoustic energy dropped significantly and ICIs became highly erratic in the latter portions of these scenes, inter-track similarity scores failed to link the remaining fragments, producing several short, unresolved acoustic clusters (four in Scene 5; three in Scene 6). Because this fragmentation prevented continuous track reconstruction, compounded by subsequent spatial ambiguities during optical validation (Sec. III.C), we conservatively excluded Scenes 5 and 6 from the comprehensive track-level performance metrics (Table II; Table S3). Nevertheless, extracting these preliminary tracks formatted most signals for subsequent spatial evaluation.

### C. Spatial localization, optical validation, and final individual attribution

To assign the spectro-temporally clustered click trains (Sec. III.B) to specific individuals, we evaluated spatial localization in parallel across all 2,308 clicks detected using the strict threshold (Sec. II.C). Applying loop closure errors and geometric residual filters retained 1,672 valid clicks across the three sessions (471 for session 081356, 870 for 085315, and 331 for 085452; Appendix Table IV). We subsequently cross-referenced this geometrically validated subset with the clustered tracks. Although we attempted spatial localization on the highly entangled Scenes 5 and 6 (comprising session 085315), unresolved vertical ambiguities forced their exclusion from final individual assignment (Sec. IV.B). Consequently, restricting our analysis to the 882 focal clicks clustered within the retained sessions (Scenes 1-4 and 7-9) yielded a final subset of 637 focal clicks (381 and 256 for sessions 081356 and 085452, respectively) available for TDOA-based bearing estimation and optical validation. Within this subset, 10 of the 637 clicks fell in a marginal regime where the tangential vector slightly exceeded the unit-norm constraint ($\|\mathbf{g}_{2D}\|^2$ in the range 1.01-1.72), corresponding to DOA near the array's resolvable cone. These clicks were retained rather than discarded (Sec. II.E), as they still satisfied the $\leq 10\ cm$ reliability threshold applied to all localizations; their effect on the aggregate concordance and error-regime results is therefore limited.

We applied probabilistic KDE to this filtered acoustic subset to extract a dominant two-dimensional spatial mode (azimuth and elevation) for each track. This statistical approach demonstrated spatial divergence between interacting individuals, even when broader spatial isopleths overlapped. For instance, in Scene 3 (Fig. 5), despite dense temporal entanglement, KDE resolved two distinct directional modes: the main speaker (Track 3) peaked at -15.1° azimuth and 78.9° elevation, while the secondary speaker (Track 4) peaked at -4.1° and 72.5°. Similarly, in Scene 8 (Fig. 3), azimuthal resolution distinguished the emitters, placing Track 8 to the right (12.4°) and Track 9 to the left (-1.4°). In single-emitter contexts (e.g., Scene 7), KDE isolated a single robust centroid (azimuth 9.7°, elevation 60.7°). These KDE-derived acoustic angles provided the necessary spatial baseline for subsequent optical validation.

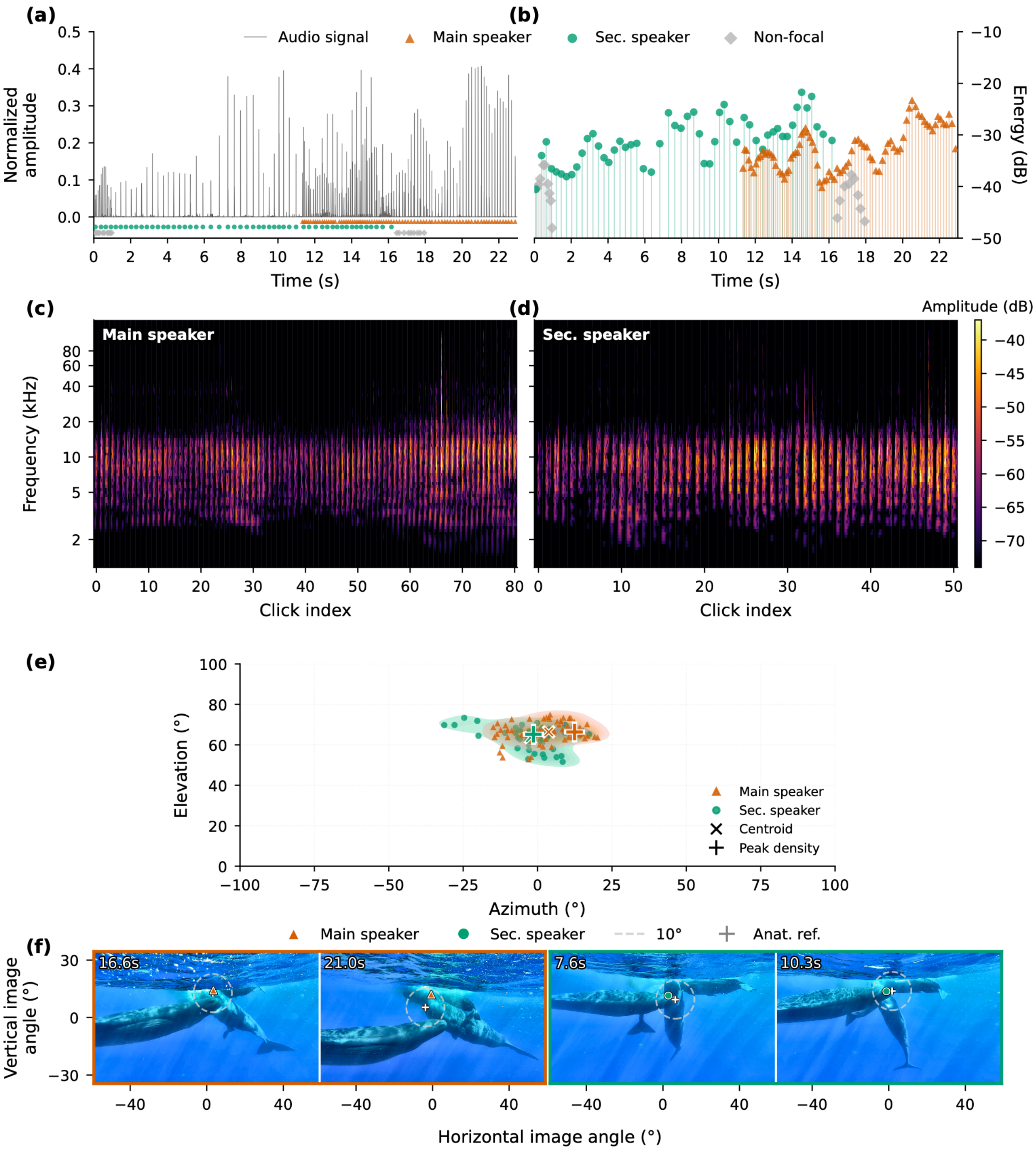

FIG. 3. (Color online). Multimodal source separation workflow across multi-emitter interaction (Scene 8). Panels display: (a) normalized acoustic waveforms with marker overlays indicating click occurrence times; (b) temporal energy profiles (dB); (c) and (d) Constant-Q Transform spectral evolution computed from pre-whitened waveforms for the main and secondary speakers, respectively (indexed by click number); (e) acoustic spatial distribution (azimuth and elevation) via Kernel Density Estimation (KDE), detailing centroids and peak densities; and (f) optical validation via projecting onto the synchronized video plane, displaying 10° spatial tolerance threshold (dashed circles) around the anatomical reference (distal end of the junk). Non-focal clicks are shown in panels (a) and (b), as they are part of the acoustic scene and contribute to the clustering performance evaluation; they are excluded from the KDE spatial analysis (e), which characterizes the spatial modes of the two focal emitters only. Orange triangles denote the main speaker (Ali, Track 8) and teal circles denote the secondary speaker (Daren, Track 9)

Subsequently, we mathematically projected the aligned angular coordinates of these 637 clicks onto the synchronized video's 2D image plane. $VCR$ (Table III) revealed disparity between horizontal and vertical axes. While horizontal visual concordance ($VCR_H$) reached 100% for several primary tracks (e.g., 1, 2, and 7), vertical concordance ($VCR_V$) dropped significantly in subsurface contexts, reaching a minimum of 21.90% for Track 5.

TABLE III. Summary of spatial projection performance, visual concordance rates horizontal and vertical ($VCR_H$ and $VCR_V$), and final individual attribution for spatially filtered focal tracks. *Note.* Tracks from Scenes 5 and 6 are excluded due to spatial ambiguities preventing definitive frontal assignment.

| Scene | Track | Role | Filtered clicks | $VCR_H$ (%) | $VCR_V$ (%) | Final ID |
|---|---|---|---|---|---|---|
| 1 | 1 | Main | 67 | 98.51 | 37.31 | Ali |
| 2 | 2 | Main | 30 | 96.67 | 63.33 | Ali |
| 3 | 3 | Main | 90 | 64.44 | 33.33 | Ali |
| 3 | 4 | Sec. | 58 | 62.07 | 43.10 | Daren |
| 4 | 5 | Main | 105 | 74.29 | 21.90 | Daren |
| 4 | 6 | Sec. | 31 | 83.87 | 25.81 | Ali |
| 7 | 7 | Single | 66 | 100 | 96.97 | Ali |
| 8 | 8 | Main | 72 | 100 | 94.44 | Ali |
| 8 | 9 | Sec. | 41 | 65.85 | 80.49 | Daren |
| 9 | 10 | Main | 57 | 42.11 | 66.67 | Daren |
| 9 | 11 | Sec. | 20 | 55.00 | 80.00 | Ali |
| Total | | | 637 | | | |

To objectively quantify the projection biases driving spatial disparities, we applied the GMM to the projection errors of the 637 spatially filtered clicks (Sec II.F). The component selection yielded $k = 4$ as the optimal solution. Although the BIC minimum occurs at $k = 5$, the four-component model demonstrated substantially higher structural stability ($ARI = 0.976 \pm 0.020$) than the five-component solution ($ARI = 0.827 \pm 0.183$), indicating the fifth component lacked reproducible structure. Physical inspection confirmed this: the additional component merely sub-divided the nominal success regime into two spatially adjacent sub-clusters, unrelated to any distinct acoustic error source. We therefore retained the four-component model as the most parsimonious and structurally robust solution, which identified four distinct operational regimes (Fig. 4, a):

Type 1 (Nominal success; n=361): A central convergence zone representing the system's optimal operating area. For instance, Scene 7 (Track 7) illustrates this regime (100% of 66 clicks within this zone) (Fig. 4, b; Mm. 1), alongside Track 8 (100% of 72) and Track 1 (91% of 67).

Type 2 (Proportional bias; n=145): A correlated, diagonal error stretching across both axes. This prominently affects Track 5 in Scene 4 (68 of 105 clicks; Fig. 4, c), accounting for the severe degradation of vertical concordance ($VCR_V$) to 21.90%.

Type 3 (Body shadowing and multipath bias; n=98): A directional spatial bias affecting oblique-posture emissions, observed in Scene 8 (28 of 41 clicks, Track 9) and Scene 9 (41 of 57 clicks, Track 10) (Fig. 4, d; Mm. 2).

Type 4 (Horizontal failure; n=33): Isolated extreme horizontal deviations. As noted in Scene 3 (Fig. 3), extreme temporal proximity induces this shift. Crucially, this error corrupts only specific temporally entangled segments, as evidenced by Track 3 (23 of 90 clicks affected), confirming that the track's overall spatial identity remains resolvable (Fig. 4, e).

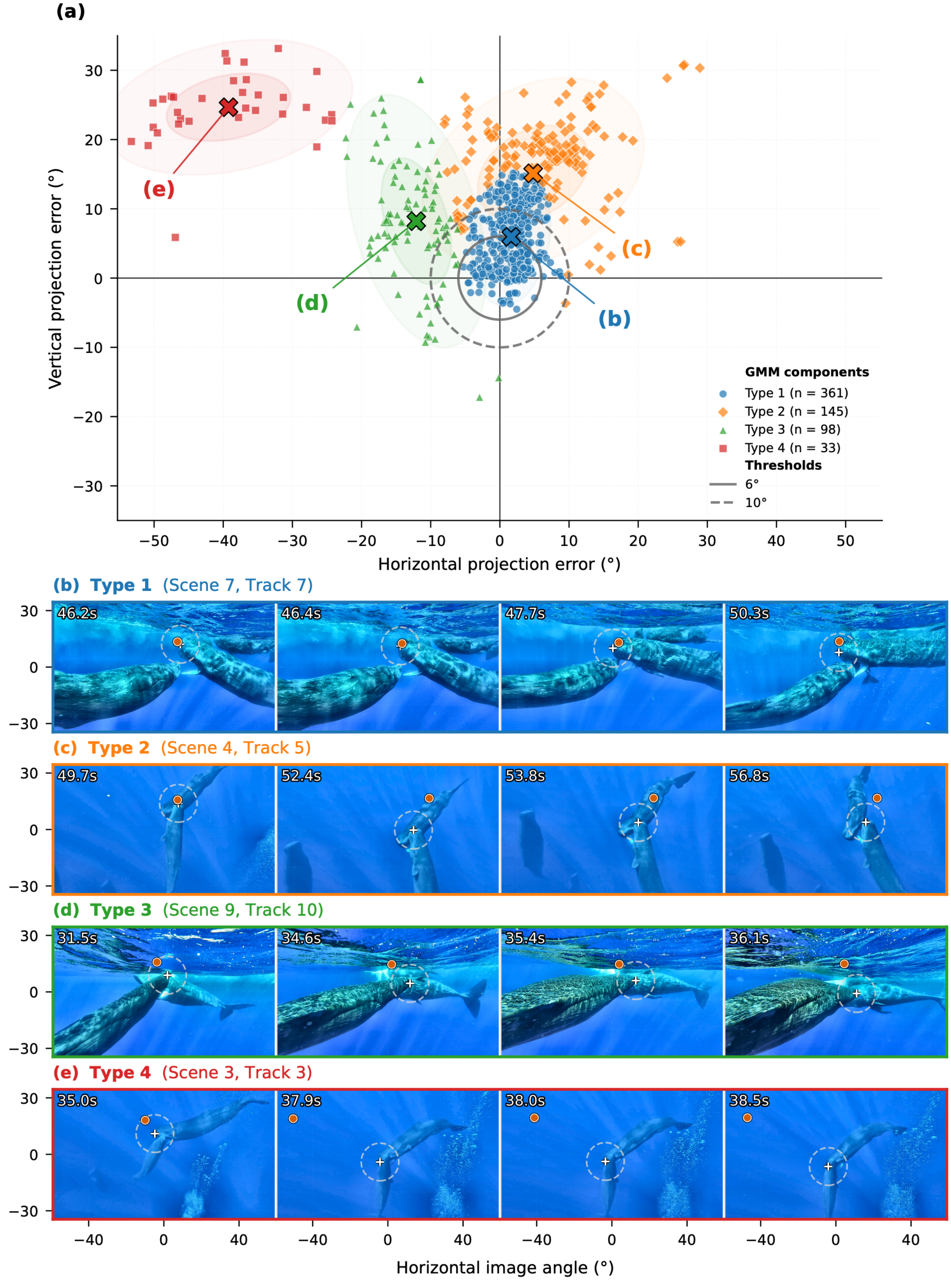

(a)
Vertical projection error (°)
Horizontal projection error (°)
(b)
(c)
(d)
(e)
GMM components
Type 1 (n = 361)
Type 2 (n = 145)
Type 3 (n = 98)
Type 4 (n = 33)
Thresholds
6°
10°
(b) Type 1 (Scene 7, Track 7)
46.2s
46.4s
47.7s
50.3s
(c) Type 2 (Scene 4, Track 5)
49.7s
52.4s
53.8s
56.8s
(d) Type 3 (Scene 9, Track 10)
31.5s
34.6s
35.4s
36.1s
(e) Type 4 (Scene 3, Track 3)
35.0s
37.9s
38.0s
38.5s
Vertical image angle (°)
Horizontal image angle (°)

FIG. 4. (Color online). Gaussian Mixture Model (GMM) clustering of multimodal spatial projection errors for 637 filtered clicks. (a) Scatter plot of horizontal vs. vertical errors (°). The origin (0°, 0°) denotes perfect concordance with the manually tracked anatomical reference (distal end of the junk). Large cross markers (x) indicate GMM component centroids; shaded ellipses represent one-standard-deviation confidence ellipses. Dashed circles indicate the 6° and 10° spatial tolerance thresholds. The GMM identifies four operational regimes: Type 1 (nominal success; n = 361; circles), Type 2 (proportional distortion; n = 145; diamonds), Type 3 (isotropic dispersion; n = 98; triangles), and Type 4 (horizontal failure; n = 33; squares). (b)-(e) Representative optical validation frames for each GMM component, mapped in the camera's image plane angular coordinate system. In each frame, the filled circle indicates the projected acoustic localization, the cross (+) marks the tracked anatomical reference, and the dashed circle represents the 10° spatial tolerance threshold. Scenes correspond to: (b) Type 1 (Scene 7, Track 7); (c) Type 2 (Scene 4, Track 5); (d) Type 3 (Scene 9, Track 10); (e) Type 4 (Scene 3, Track 3).

Two multimedia files illustrate the click assignment procedure and the associated spatial error regimes. These videos overlay projected acoustic localizations onto synchronized frames for both focal individuals across retained scenes.

Mm. 1. Optical validation sequence for Scene 7 (single emitter, Ali). Orange triangle markers denote projected acoustic localizations within the image plane angular coordinate system (118° x 69° FOV). Track 7 demonstrates nominal Type 1 (Fig. 4, a,b) operation with 100% horizontal and 96.97% vertical concordance (Table III, Track 7).

Mm. 2. Optical validation sequence for Scene 8 (two emitters, Ali and Daren). Acoustic localizations are color-coded by emitter: orange triangles for the main speaker (Track 8, Ali) and teal circles for the secondary speaker (Track 9, Daren). Track 9 illustrates the Type 3 error regime (Fig. 4, a,d), demonstrating the directional spatial bias induced by body shadowing during oblique postures, with 65.85% horizontal and 80.49% vertical concordance (Table III, Track 9).

The complete set of rendered optical-validation videos (MP4) for all nine focal interaction scenes, including the unresolved Scenes 5 and 6, is available in the data repository (see Data Availability); Mm. 1 and Mm. 2 present two representative examples.

Error types across recording sessions reveals that Type 2 errors (proportional distortion; n=145) predominantly concentrated in subsurface interaction contexts from session 081356 (e.g., Track 5, Scene 4; 68 of 105 clicks). Conversely, surface interactions from session 085452 predominantly cluster within the nominal success regime (Type 1). We further discuss this session-dependent vertical projection error pattern in Sec. IV.C.

We empirically compared visual concordance at 6° and 10° spatial tolerance thresholds across the dataset. Horizontal concordance was largely insensitive to this choice, remaining high at both thresholds. The vertical axis, by contrast, was markedly affected: the 6° threshold substantially degraded vertical concordance in subsurface contexts, whereas the 10° threshold preserved it. We therefore adopted the 10° threshold for all reported concordance rates.

Although the GMM characterized these operational regimes, extreme behavioral configurations in Scenes 5 and 6 exposed the pipeline's spatial limits. In Scene 5, visual inspection confirmed the subjects' extreme physical proximity and exact vertical alignment. Consequently, the two primary acoustic sequences (55 and 37 clicks) exhibited intertwined KDE distributions, revealing overlapping spatial isopleths with nearly identical centroids that prevented separation. Similarly, focal subjects in Scene 6 moved constantly with vertically stacked acoustic emission centers (melons). Furthermore, as subjects moved away from the array's central axis, short track fragments failed to localize robustly due to off-axis camera distortion and signal degradation.

Because the subjects in these saturated scenes aligned nearly horizontally, azimuthal tracking proved insufficient. Relying on the array's less precise elevation estimates, combined with acoustic fragmentation, prevented unambiguous frontal attribution of these sequences to either Ali or Daren, justifying their exclusion from Table III.

The optical modality compensated for these acoustic spatial illusions. Robust horizontal visual concordance ($VCR_H$) resolved front/back ambiguities, guaranteeing unambiguous attribution for most of the dataset. Consequently, visual validation confirmed individual acoustic dominance (Table III): Ali was the primary or sole emitter in Scenes 1, 2, 7, and 8 (Track 8), as well as the dominant emitter in Scene 3 (Track 3) and the secondary emitter in Scenes 4 and 9. Conversely, Daren dominated Scenes 4 (Track 5) and 9 (Track 10), while acting as the secondary emitter in Scenes 3 and 8. These assignments established the individual origin of the click trains produced during the analyzed vocal sparring interactions.

## IV. DISCUSSION

### A. Overcoming near-field separation and attribution limits

As established, standard TDOA tracking and multipath-based separation are effective for deep-diving sperm whales (Giraudet and Glotin, 2006; Thode, 2004, 2005) but fail in the upper 20 m of the pelagic environment (Baggenstoss, 2011; Gubnitsky et al., 2025; Zimmer, 2011). This physical constraint is compounded by hardware limitations inherent to compact heterogeneous arrays. While spectral and temporal extractions utilize a dedicated reference channel (SQ26) to avoid cross-sensor biases, spatial tracking relies on the entire array. The distinct transient responses and phase disparities between the omnidirectional SQ26 and the broadband C75 sensors degrade cross-correlation accuracy, rendering pure spatial tracking vulnerable to spurious correlation peaks.

Beyond the demographic indistinguishability discussed previously (Barile et al., 2024; Caruso et al., 2015), IPI-based separation in near-surface contexts fails because the internal pulse structures undergo physical degradation. In the immediate surface layer (0-2 m), pulse smearing and destructive phase cancellation via the LME (Pereira et al., 2020; Urick, 1983) destroy the transient definition required for IPI extraction. While deep-diving studies exploit or reject delayed surface reflections (Baggenstoss, 2011; Nosal and Neil Frazer, 2006; Thode, 2004; Zimmer, 2011), sub-surface emissions (2-20 m) introduce high-amplitude multipath arrivals whose reflection delay exceeds the internal pulse duration. If processed through standard automated pipelines, these delayed echoes risk being conflated with secondary internal pulses, generating artifactual, bimodal IPI distributions (Sec. III.B) that mimic biological signatures and create spatial aberrations (i.e., "ghost whales").

Because IPI extraction is unviable, our algorithm couples ICI rhythmic predictions with spectral consistency tracking via the CQT. Applying CQT overcomes the fixed time-frequency resolution limits of standard STFTs. Crucially, integrating a first-order discrete-time derivative operator (Eq. (3)) as a pre-whitening filter flattens the $1/f$ spectral tilt, rebalancing the spectral envelope to expose the abrupt biological click onsets and thereby improving intra-individual similarity across a broad complexity gradient. Although CQT inherently loses discriminative power during perfect temporal superpositions, where the dominant emitter's energy deforms the weaker track's CQT patches (Fig. 3), our multimodal framework relies on optical validation to resolve these residual acoustic ambiguities.

Critically, the exclusion of Scenes 5 and 6 from the primary performance tables (Tables II and III) reflects a specific geometric limitation of the near-planar array, not a fundamental failure of the acoustic framework. During these two scenes, acoustic clustering successfully assigned 83.6% and 80.0% of validated clicks, respectively, maintaining strong spectral coherence (Table S3). Subsequent spatial validation alone drove their exclusion: the subjects' near-identical vertical alignment produced overlapping spatial isopleths with nearly identical KDE centroids. This prevented unambiguous frontal attribution, highlighting a geometric constraint of the near-planar array (Sec. IV.B). However, reporting the full clustering metrics in Supplementary Table S3 confirms that the spectro-temporal acoustic stage remains functional even at these boundary conditions.

### B. Physical and algorithmic origins of spatial errors (GMM interpretation)

Beyond individual tracking, the GMM unmixes spatial projection errors into distinct operational regimes (Fig. 4). Intersecting these clusters with kinematic metadata reveals that spatial errors originate primarily from TDOA degradation under complex acoustic conditions, quantifying the physical limitations of deploying a compact, near-planar three-hydrophone array in reflective near-surface environments. Specifically, the significant degradation in vertical $VCR_V$ (dropping to 21.9% for specific tracks; Table III) exposes a systematic elevation bias, which is further investigated in Sec. IV.C.

Because the heterogeneous sensor configuration (two SQ26s and one broadband C75) is constrained to a tightly inclined plane, the vertical baseline provides inherently lower spatial resolution than the horizontal axis. Surface multipath propagation exacerbates this ambiguity. Consequently, when individuals align vertically, the system faces an ill-posed localization problem driven by the array's restricted volumetric rank (Spiesberger and Wahlberg, 2002).

The 4-component GMM (Fig. 4) isolates three error regimes beyond the nominal success (Type 1, characterized in Sec. III.C):

Geometric non-linearity (Type 2; n=145): This diagonal, proportional error highlights a projection constraint at the periphery of the FOV. Mapping TDOA hyperbolas onto rectilinear optical pixels induces a proportional distortion. The array's near-planar geometry compounds this effect by compressing elevation tracking, a direct mathematical consequence of this same geometric constraint that compromises orthogonal resolution (Zimmer and Troiano, 2024).

Body shadowing and multipath (Type 3, n = 98): This directional spatial bias predominantly stems from oblique animal postures. Because the primary sperm whale acoustic pulse ($P_1$) is forward-directional, "tail-on" orientations physically masks the phonic lips, attenuating the direct wavefront. The cross-correlation algorithm consequently locks onto diffracted lateral leaks or surface-reflected components, generating directionally biased pointing vectors.

Correlation phantoms (Type 4, n = 33): Extreme horizontal failures occur during temporal click collision between concurrent emitters. These overlaps fool the generalized cross-correlation pipeline, inducing systematic horizontal displacements termed "correlation phantoms", which represent a fundamental algorithmic limit of generalized cross-correlation in the absence of sufficient temporal separation. However, because this regime corrupts only specific temporally entangled segments (e.g., 23 of 90 clicks in Track 3; Fig. 5), overall spatial identity remains resolvable (Fig. 4).

Beyond these algorithmic limitations, physical obstacles pose distinct challenges. The array's near-planar geometry induces "elevation blinding" (Spiesberger and Wahlberg, 2002; Wahlberg et al., 2001) during vertical subject alignment. Separately, signal saturation at close range and anatomical body shadowing introduce signal distortions. Because acoustic metrics alone cannot resolve these physical phenomena (Sec. III.C), multimodal optical validation is necessary.

### C. Physical origin of the systematic elevation offset: array geometry and the $P_0/P_1$ hypothesis

We anchored the visual ground truth, the image origin (0, 0), to the distal end of the junk. As the primary radiating surface of the sperm whale's directional biosonar, the junk represents the most stable macroscopic landmark for frame-by-frame video tracking. However, even during nominal Type 1 operation (Fig. 4), GMM analysis reveals a systematic positive vertical offset (+5° to +8°) between the acoustic elevation and this visual reference.

Mechanistically, this offset explains the axis-dependent effect of the spatial tolerance threshold. Because this +5° to +8° shift structurally displaces localized clicks beyond a 6° vertical boundary independently of the system's underlying accuracy, a 6° threshold penalizes otherwise correctly attributed clicks. Azimuthal estimates, unaffected by this vertical bias, remain stable. Consequently, the 10° threshold represents the minimum value required to encompass this systematic offset. Spanning only 8.5% and 14.5% of the horizontal and vertical FOV, respectively, it remains the spatial specificity necessary for individual-level discrimination. Because this threshold was calibrated on the present dataset, the reported $VCR$ reflect the pipeline's internal spatial consistency rather than an independent external validation.

Two non-exclusive interpretations explain this systematic offset. First, the near-planar geometry of the three-hydrophone sub-array likely induces a structural vertical bias in the image plane (Sec. IV.B). The limited vertical aperture and heterogeneous sensor configuration (SQ26 vs. C75) degrade TDOA-derived elevation estimates, potentially generating a systematic positive shift under surface multipath conditions. Second, the array may preferentially track the initial generating pulse ($P_0$) at the phonic lips rather than the main beam ($P_1$) emitted at the distal end of the junk, aligning with the widely accepted multipulse sound production mechanism (Møhl et al., 2003; Zimmer et al., 2005). Three physical mechanisms support this anatomical interpretation:

1. Chronological locking: our signal processing pipeline implements the non-linear TKEO (Eq. (1)) to identify the precise onset of the click envelope (Sec. II.C). Physically, $P_0$ is generated first at the phonic lips (Møhl et al., 2003). While a fraction of this energy radiates directly into the water as $P_0$, the remainder propagates backward through the spermaceti organ, reflects off the frontal sac, and exits through the junk as the delayed $P_1$ beam. Because $P_0$ arrives first, adaptive thresholding naturally triggers on this transient, forcing the TDOA cross-correlation to lock onto the $P_0$ phase center.

2. Off-axis directivity loss: the primary $P_1$ beam is forward-directed. Given the dynamic, tactile, and oblique postures adopted during surface socialization (Fig. 4, d), the hydrophone array is rarely aligned within the optimal $P_1$ acoustic axis. In these off-axis configurations, the high-frequency energy of $P_1$ collapses. Consequently, the more omnidirectional, low-frequency leakage of the initial $P_0$ pulse emerges as the most coherent wavefront available for TDOA tracking (Laplanche et al., 2006).

3. Wavefront integrity: because the $P_0$ pulse radiates directly from the phonic lips without propagating through the highly refractive, dissipative spermaceti and junk complex, its wavefront preserves high structural integrity. This minimal dispersion maximizes the peak sharpness of the generalized cross-correlation function (Sec. II.E), providing a more stable mathematical convergence for $P_0$ than for the heavily diffracted and multipath-corrupted $P_1$ pulse.

Furthermore, the session-dependent distribution of spatial errors, predominantly concentrated in recording session 081356 (subsurface interaction contexts; Sec. III.C), is consistent with both interpretations. Geometrically, subsurface interactions involve complex multipath conditions that exacerbate TDOA elevation estimation errors. Anatomically, subsurface interactions may produce clearer temporal separation between $P_0$ and $P_1$, reinforcing TKEO locking on the first-arriving $P_0$ wavefront.

The current dataset cannot conclusively disentangle geometric and anatomical contributions to this offset, as both predict a systematic positive vertical shift of similar magnitude. Rigorously distinguishing them requires either a volumetric array with sufficient vertical aperture or a controlled experiment using an acoustic source at known varying depths. Therefore, we regard the $P_0/P_1$ interpretation as a physically motivated hypothesis, rather than a demonstrated result. Nonetheless, both interpretations converge on the same practical implication: future array deployments must improve vertical aperture to resolve this ambiguity. If the anatomical hypothesis is confirmed, such systems could achieve direct optical validation of acoustic phase center tracking in freely socializing odontocetes.

### D. Spectro-temporal variability and behavioral implications

Assigning atypical, short-range social signals represents a major contribution of this framework. By attributing these overlapping click trains to identified individuals, our workflow establishes that the emissions co-occurring with vocal sparring are produced by the interacting whales rather than by surrounding conspecifics, confirming, for this dyad, that these click trains are a genuine component of the behavior and not merely concurrent background activity. This association nonetheless rests on a single dyad and one behavioral context, and generalizing it will require larger, multi-individual datasets.

These "vocal sparring" emissions feature broad bandwidths, variable rhythms, and energy modulations. Applying the CQT overcomes the low-frequency spectral blurring of standard STFTs, indicating that observed lower fluctuations below 5 kHz are likely mechanical rather than biological. These spectral signatures results from complex multipath propagation, body masking, and dynamic off-axis effects (Møhl et al., 2003; Teloni et al., 2007). Because the biosonar beam is directional (Jensen et al., 2018), simple kinematic rotations dynamically filter the received spectral envelope.

Depth-dependent observations (Fig. 2) illustrate this physical reality. Interactions within 1 m of the surface manifest as substantial global attenuation via a classic LME. While deep-diving studies exploit predictable multipath delays to inter depth (Nosal and Neil Frazer, 2006; Pereira et al., 2020; Thode, 2004), our surface-layer array geometry compresses these delays, transforming surface reflections from a tracking metric into destructive interference. Conversely, at 15 m, extended delays shift destructive notches across the spectrum, systematically restoring low-frequency energy. Future behavioral signal validation will thus require blind deconvolution to strip away this environmental multipath response.

These mechanical and environmental factors must be contextualized against recent hypotheses regarding voluntary vocal control, such as vowel coarticulation (Beguš et al., 2025; Sharma et al., 2024) and shifts between laryngeal-like vocal registers (Madsen et al., 2023). Because recent findings predominantly rely on animal-borne biologgers (D-tags) that insulate the recorded signal from propagation distortions, we advocate strict ethological caution when interpreting off-body array recordings. While rhythmic modulations during vocal sparring suggest intentional dialogue, apparent spectral modulations, such as those sometimes attributed to coda coarticulation, can, in array recordings, often be pure propagation or “pass-by” kinematic artifacts rather than active vocal control.

Here, our multimodal architecture provides a critical advantage over visually blind arrays. By supplying optical ground truth of subject orientation and jaw kinematics, the system explicitly disentangles physical off-axis propagation effects from genuine vocal modulation. While these findings are a technical proof-of-concept, they underscore that validating complex phonology in multi-emitter interactions requires joint kinematic-acoustic analysis on larger datasets to definitively isolate active vocal control from severe physical artifacts documented herein.

## V. CONCLUSION

This work presents a multimodal audio-visual workflow that resolves the methodological bottleneck of acoustic diarization during dense, near-surface socialization in sperm whales. By integrating spectro-temporal tracking with optical validation, this non-invasive approach successfully assigned 11 distinct, intertwined click trains totaling 882 clicks to two specific immature males (Ali and Daren) engaging in vocal sparring. Despite the constraints of operating with a compact near-surface array, the pipeline proved effective, where purely acoustic TDOA spatial estimates failed due to mechanical biases, visual concordance provided the critical ground truth to resolve ambiguities. Designed to generalize beyond this specific dyad, this framework could reliably track the underrepresented juvenile male demographic or differentiate individuals with indistinguishable IPIs.

From an ethological perspective, decoding these interactions expands the understanding of marine mammal social intelligence beyond traditional coda research. The vocal sparring signals isolated here, superficially resembling foraging echolocation clicks yet emitted during tactile social exchanges, reveal a previously undescribed dimension of non-coda acoustic communication. Applying this framework to larger datasets will be essential to assess whether these signals conceal context-dependent modulations, correlating detailed body kinematics with reliably assigned emissions to provide a holistic view of socio-acoustic networks and vocal ontogeny.

Crucially, this study achieves accurate attribution rather than pristine signal restoration. Because the raw waveforms remain heavily distorted by dynamic off-axis emissions and surface multipath interference, any future structural characterization requires rigorous signal reconstruction. Rather than relying solely on blind deconvolution, coupling our optical kinematic data with established anatomical acoustic models (e.g., Laplanche et al., 2006) opens promising avenues for informed signal deconvolution. By capturing the exact head orientation required to theoretically invert these models, the synchronized optical modality remains an imperative. Moving forward, this multimodal foundation provides the most direct mechanism to mathematically strip away physical kinematic artifacts, thereby isolating genuine biological modulation and rigorously decoding the conversational complexity of free-ranging sperm whales.

## SUPPLEMENTARY MATERIAL

See supplementary material for a description of the OPALE array hardware specifications and geometry (S1) ; the acoustic calibration and detection parameters (S2); the ROC curve analysis and threshold optimization (S3); the spectro-temporal tracking and clustering parameters (S4); the automatic transient detection and manual validation performance (Table S1); the inter-click interval (ICI) statistics and coefficient of variation (CV) for all twelve interaction scenes (Table S2); the partial acoustic clustering metrics for the excluded Scenes 5 and 6 (Table S3); and the multimodal source separation workflows for the remaining interaction scenes (1, 2, 4, 7, 9) (Figs. S1-S5).

### S1. OPALE array hardware specifications and geometry

This section details the hardware specifications and the exact spatial configuration of the OPALE array utilized during the 2023 deployment for multimodal data acquisition.

#### Acoustic sensors and acquisition hardware:

The functional three-hydrophone sub-array comprises heterogeneous sensors manufactured by Cetacean Research Technology (Seattle, WA, USA).

Although the complete OPALE array integrates five hydrophones, hardware constraints during the 2023 deployment limited the system to a functional sub-array of three heterogeneous sensors (Cetacean Research Technology, Seattle, WA, USA). Channels 0 and 1 ($H_0$ and $H_1$), which define the primary horizontal axis of the array, are equipped with SQ26-H1B hydrophones. These sensors operate within a bandwidth of 20 Hz to 45 kHz and maintain omnidirectional sensitivity for frequency below 10 kHz. Channel 4 ($H_2$), positioned lower and forward to establish the inclined detection plane necessary for three-dimensional spatial resolution, is equipped with a C75 broadband hydrophone. This sensor provides a linear response (± 3 dB) across a frequency range of 10 Hz to 170 kHz.

Acoustic signals were digitized and stored using the on-board Qualilife Highblue (QHB) acquisition system (Barchasz et al., 2020), operating at a sampling rate of 256 kHz with a 24-bit resolution to fully capture the high-frequency transient nature of the sperm whale clicks.

**Array geometry and coordinate matrix:**

The precise spatial configuration of the hydrophones is a fundamental parameter for the Time Difference of Arrival (TDOA) spatial localization algorithms described in the main text. In the local coordinate system of the array (x-axis forwards, y-axis lateral left, z-axis upward), the exact positions of the three sensors (expressed in centimeters) are defined by the coordinate matrix H:

$$H = \begin{pmatrix} 0 & 0 & 0 \\ 0 & 51.3 & 0 \\ 21.5 & 25.65 & -30.29 \end{pmatrix}$$

**S2. Acoustic calibration and detection parameters**

Calibration: hardware sensitivities were applied at -169 dB re 1V/µPa for the SQ26 sensors and -180 dB re 1V/µPa for the C75 sensor. A linear amplitude factor of 3 and a polarity inversion and a polarity inversion were applied specifically to the C75 channel prior to TDOA calculation.

TKEO normalization: the TKEO trace was adaptively normalized using a sliding-window local median, then smoothed with a recursive leaky integrator. The window length and the integrator time constant were tuned empirically to match the temporal scale of the click envelopes while suppressing residual background fluctuations (Data Availability).

Envelope tracking: the adaptive envelope tracking filter was implemented over a 15-s sliding window, explicitly excluding any signals falling 15 dB below the local 90th percentile of click energy.

**S3. ROC curve analysis and threshold optimization**

Detection thresholds were optimized on a single representative, manually annotated recording (file 085452; 94.56 s, 510 validated clicks) and uniformly applied across all sessions, as single-to-noise ratios and environmental conditions were broadly comparable across the deployment.

A windowed Receiver Operating Characteristic (ROC) analysis was performed on the TKEO-processed signal (Fawcett, 2006; Zimmer, 2011). The signal was segmented into 4-ms window (1024 samples at 256 kHz). To prevent penalizing transient clicks spanning adjacent frames, an 8-ms ground-truth tolerance was implemented wherein an annotated click validated both its containing window and the subsequent one.

Thresholds were empirically varied from 0 to 200 in increments of 5, classifying each window as signal or noise against the manual ground truth. This analysis indicated that a uniform array-wide threshold would degrade overall performance: the broadband C75 sensor tolerates a lower threshold, whereas the SQ26 sensors require a higher threshold to reject sensor-specific fluctuations. Consequently, the two-stage strategy applied permissive threshold (40 for SQ26; 30 for C75) to maximize track continuity for clustering, and strict thresholds (80 for SQ26; 55 for C75) to ensure a near-zero constant false alarm rate (CFAR) for TDOA localization, thereby preventing spatial aberrations.

### S4. Spectro-temporal tracking and clustering parameters

CQT and patch extraction: the CQT was computed across a frequency range of 1.2 kHz to 80 kHz with a resolution of 12 bins per octave, yielding a Quality factor ($Q$) of approximately 16.8. Implementation was optimized via an overlap-add convolution procedure in the frequency domain, incorporating rigorous group delay correction to ensure temporal alignment across all frequency bands. To precisely align spectral features, each click detected by the TKEO stage was finely realigned to its absolute energy peak within a 4-ms search window. Subsequently, a symmetric 9-ms time-frequency patch was extracted, converted to decibels, flattened, and Z-score normalized per frequency band to isolate the structural profile from absolute propagation losses.

Clustering and tracking algorithm: the adaptive tracking algorithm assigned candidate clicks based on a unified similarity metric combining cosine distance and Pearson correlation. To maintain robustness against slight temporal jitter, this composite score was evaluated across three relative temporal alignments. The predictive rhythmic constraint utilized a temporal acceptance window based on the sliding median of recent track ICIs. A ±25% physiological tolerance was applied to this window; deviations incurred a quadratic penalty to the overall similarity score, and candidates falling beyond this ±25% boundary were definitively rejected, ensuring tight rhythmic adherence during dense temporal overlaps.

TABLE S1. Automatic transient detection performance and manual ground-truth validation across the three recording sessions. Consistent with a highly sensitive detection strategy, the pipeline yielded a high recall rate (96.16%) at the expense of precision. False positives (primarily surface noise) were systematically discarded during manual validation to ensure a pristine dataset for spatial processing.

| Session | Total detections | Ground truth | True positives (detected) | False positives | False negatives (missed) |
|---|---|---|---|---|---|
| 081356 | 1,158 | 850 | 835 | 323 | 15 |
| 085315 | 1,445 | 1,295 | 1,261 | 184 | 34 |
| 085452 | 558 | 510 | 457 | 101 | 53 |
| Total | 3,161 | 2,655 | 2,553 | 608 | 102 |
| Metrics | | | Recall: 96.16% | Precision: 80.77% | Miss rate: 3.84% |

TABLE S2. Inter-click interval (ICI) statistics and coefficient of variation (CV) for all twelve interaction scenes. Scenes with $CV \geq 0.5$ were classified as multi-emitter socio-acoustic environments.

| Scene | Opale file | Analysis window (s) | Nb clicks | ICI mean (s) | ICI std (s) | CV |
|---|---|---|---|---|---|---|
| 1 | 081356 | 11.12 - 29.46 | 104 | 0.178 | 0.087 | 0.487 |
| 2 | 081356 | 85.05 - 96.36 | 70 | 0.164 | 0.081 | 0.494 |
| 3 | 081356 | 33.13 - 46.05 | 192 | 0.068 | 0.038 | 0.567 |
| 4 | 081356 | 48.24 - 69.50 | 289 | 0.074 | 0.053 | 0.720 |
| 5 | 085315 | 86.49 - 96.47 | 123 | 0.082 | 0.048 | 0.581 |
| 6 | 085315 | 0.17 - 23.86 | 188 | 0.127 | 0.076 | 0.603 |
| 7 | 085452 | 43.77 - 52.20 | 82 | 0.105 | 0.016 | 0.152 |
| 8 | 085452 | 0.08 - 22.88 | 157 | 0.146 | 0.102 | 0.697 |
| 9 | 085452 | 25.10 - 40.85 | 163 | 0.097 | 0.068 | 0.697 |
| 10 | 085315 | 28.04 - 37.33 | 289 | 0.032 | 0.027 | 0.849 |
| 11 | 085315 | 38.88 - 80.07 | 653 | 0.049 | 0.046 | 0.936 |
| 12 | 085452 | 54.36 – 62.22 | 56 | - | - | -[a] |

[a] Scene 12: CV not reported owing to incomplete detection (most of the 53 undetected clicks in session 085452 occur in this scene), which would therefore not reflect the true emission rhythm.

TABLE S3. Partial acoustic clustering metrics for the excluded Scenes 5 and 6 (session 085315). Despite successful acoustic assignment rates and strong spectral coherence, these scenes were excluded from final spatial localization due to severe geometric ambiguities preventing unambiguous frontal attribution. In the final column, spectral coherence is reported as discrete median Constant-Q Transform (CQT) values for the two sub-tracks in Scene 5, and as a range for the four sub-tracks in Scene 6.

| Scene | Total clicks | Tracks extracted | Clicks assigned | Spectral coherence (median CQT) of sub-tracks |
|---|---|---|---|---|
| 5 | 110 | 2 sub-tracks + 4 unresolved | 92/110 (83.6%) | 0.896/0.822 |
| 6 | 185 | 4 sub-tracks + 3 unresolved | 148/185 (80.0%) | 0.71-0.87 |

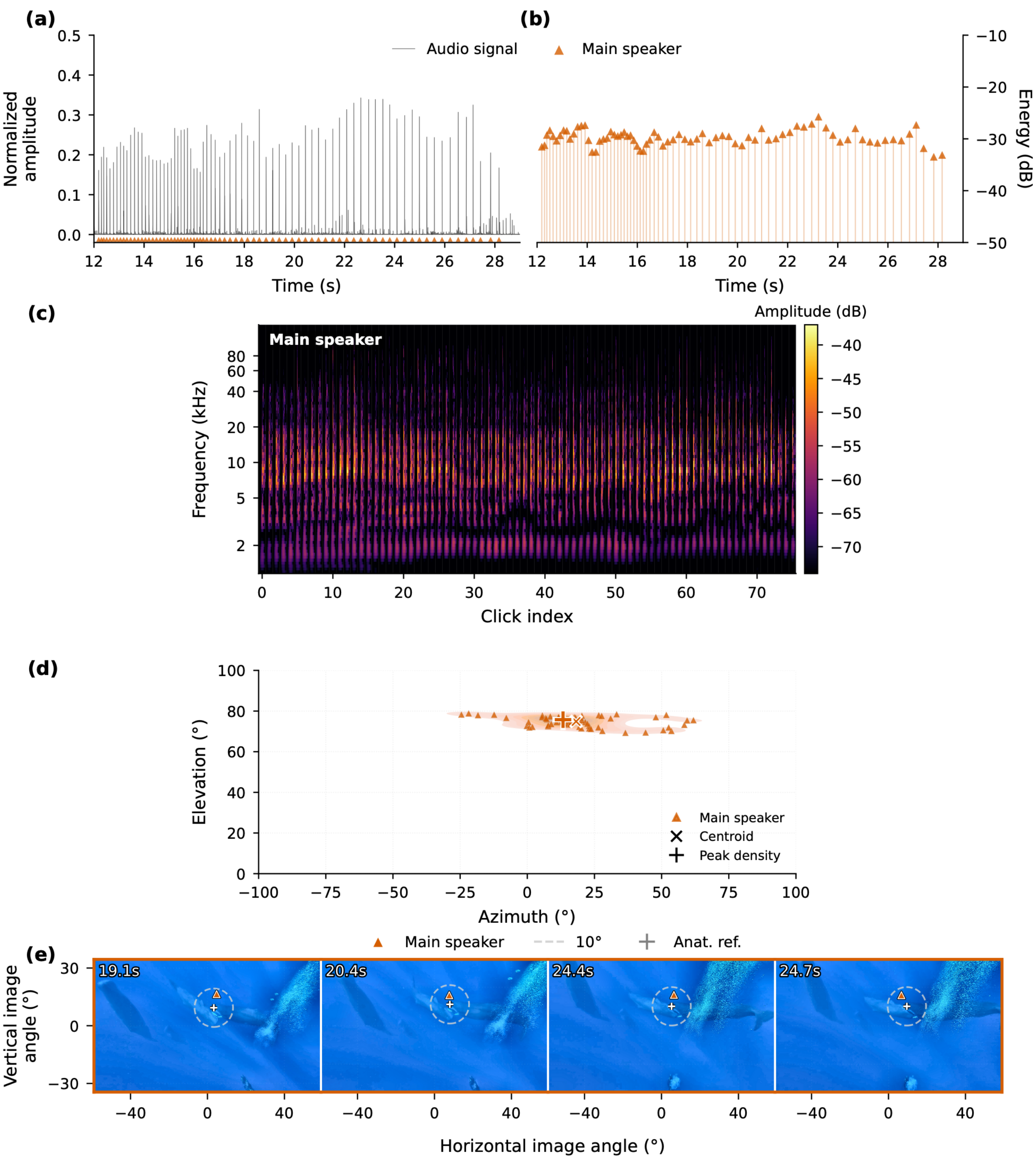

FIG. S1. (Color online). Multimodal source separation workflow across a moderately complex environment (Scene 1). Panels display: (a) normalized acoustic waveforms with marker overlays indicating click occurrence times; (b) temporal energy profiles (dB); (c) Constant-Q Transform (CQT) spectral evolution computed from pre-whitened waveforms for main speaker; (d) acoustic spatial distribution (azimuth and elevation) via Kernel Density Estimation (KDE), detailing centroids and peak densities; and (e) optical validation via projecting onto the synchronized video plane, displaying the 10° spatial tolerance threshold (dashed circles) around the anatomical reference (distal end of the junk). Orange triangles denote the main speaker (Ali, Track 1).

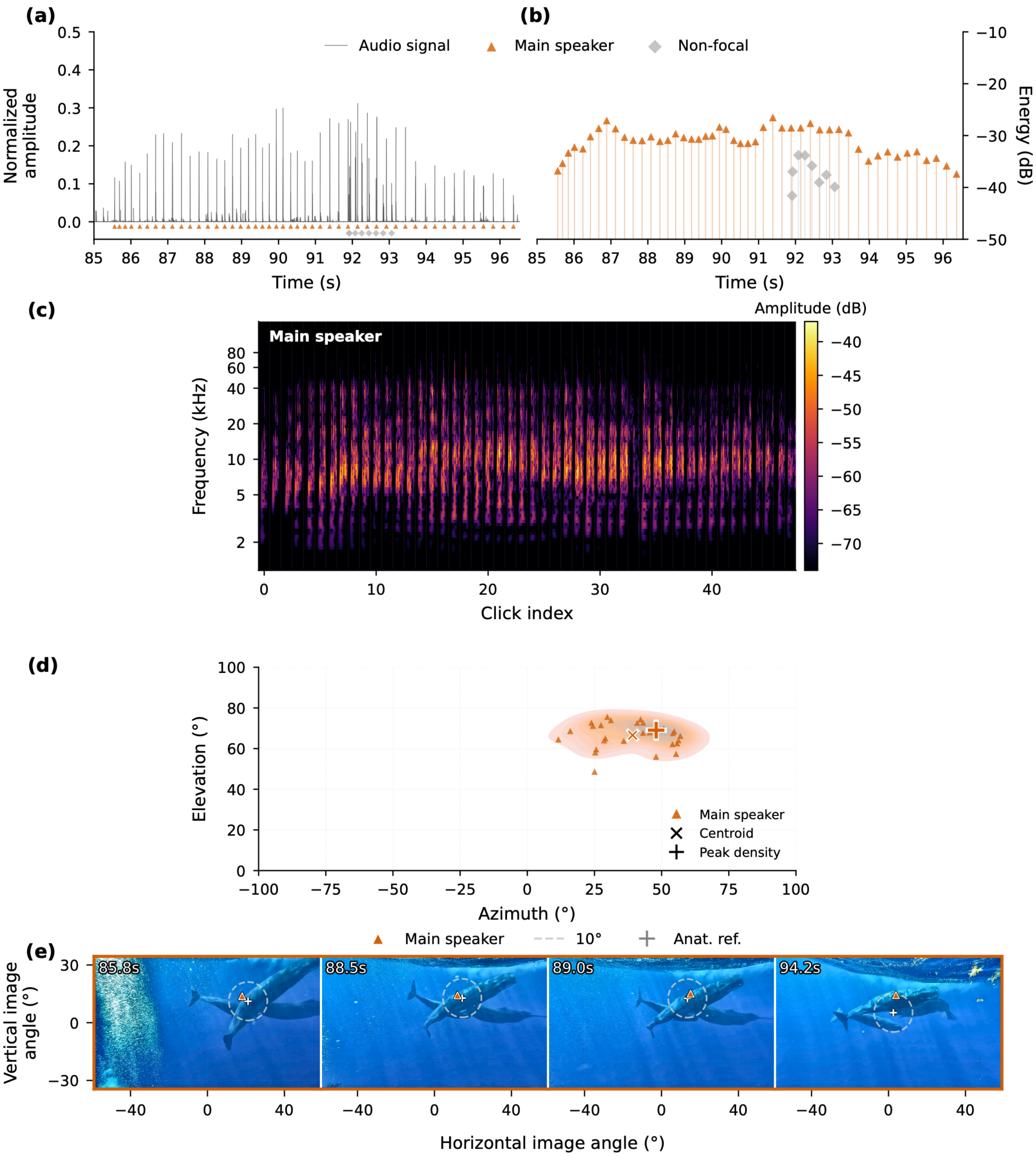

FIG. S2. (Color online). Multimodal source separation workflow across a moderately complex environment (Scene 2). Panels display: (a) normalized acoustic waveforms with marker overlays indicating click occurrence times; (b) temporal energy profiles (dB); (c) CQT spectral evolution computed from pre-whitened waveforms for main speaker; (d) acoustic spatial distribution (azimuth and elevation) via KDE, detailing centroids and peak densities; and (e) optical validation via projecting onto the synchronized video plane, displaying the 10° spatial tolerance threshold (dashed circles) around the anatomical reference (distal end of the junk). Orange triangles denote the main speaker (Ali, Track 2).

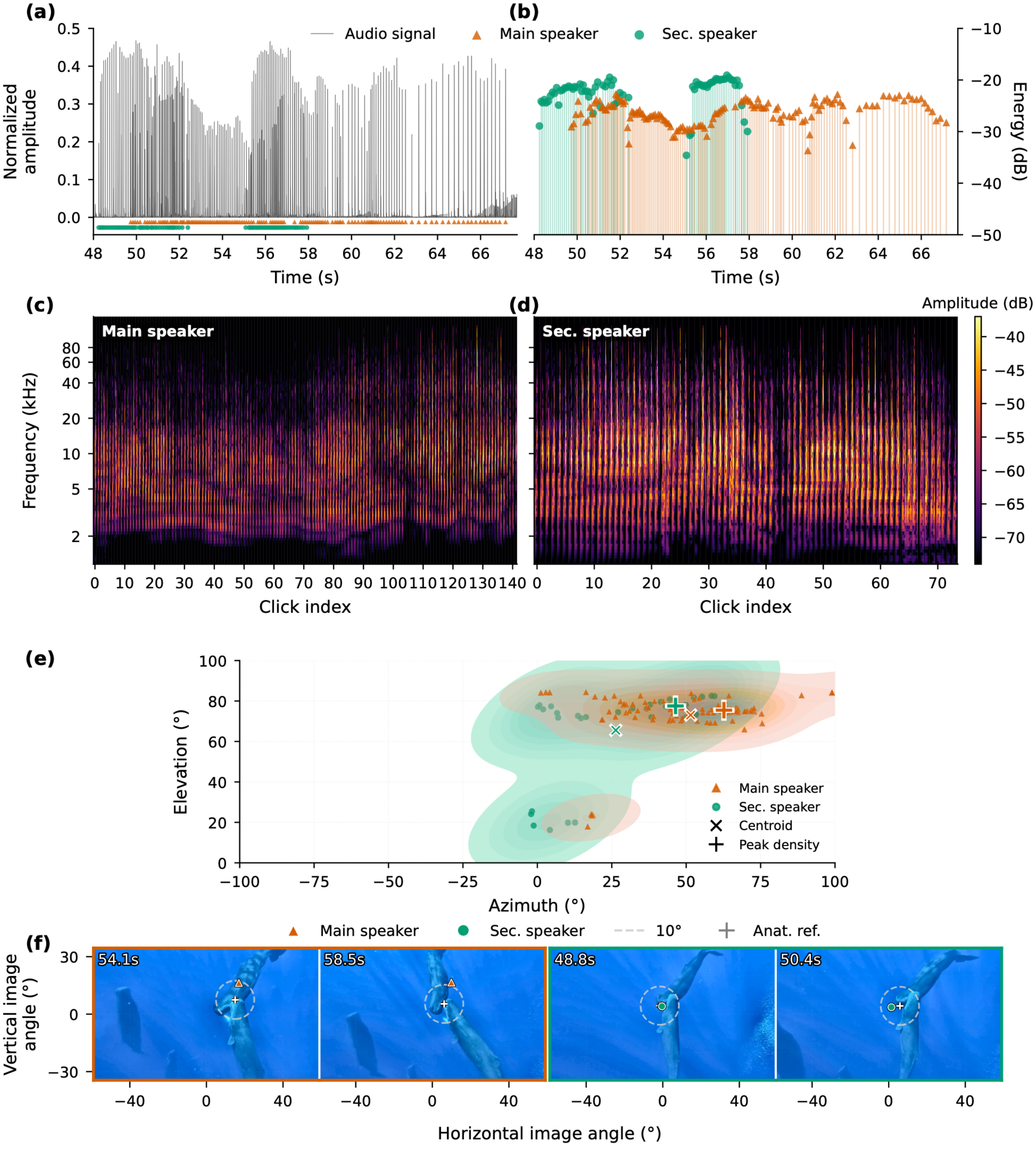

FIG. S3. (Color online). Multimodal source separation workflow across multi-emitter interaction (Scene 4). Panels display: (a) normalized acoustic waveforms with marker overlays indicating click occurrence times; (b) temporal energy profiles (dB); (c) and (d) CQT spectral evolution computed from pre-whitened waveforms for main and secondary speakers; (e) acoustic spatial distribution (azimuth and elevation) via KDE, detailing centroids and peak densities; and (f) optical validation via projecting onto the synchronized video plane, displaying the 10° spatial tolerance threshold (dashed circles) around the anatomical reference (distal end of the junk). Orange triangles denote the main speaker (Daren, Track 5); teal circles denote the secondary speaker (Ali, Track 6).

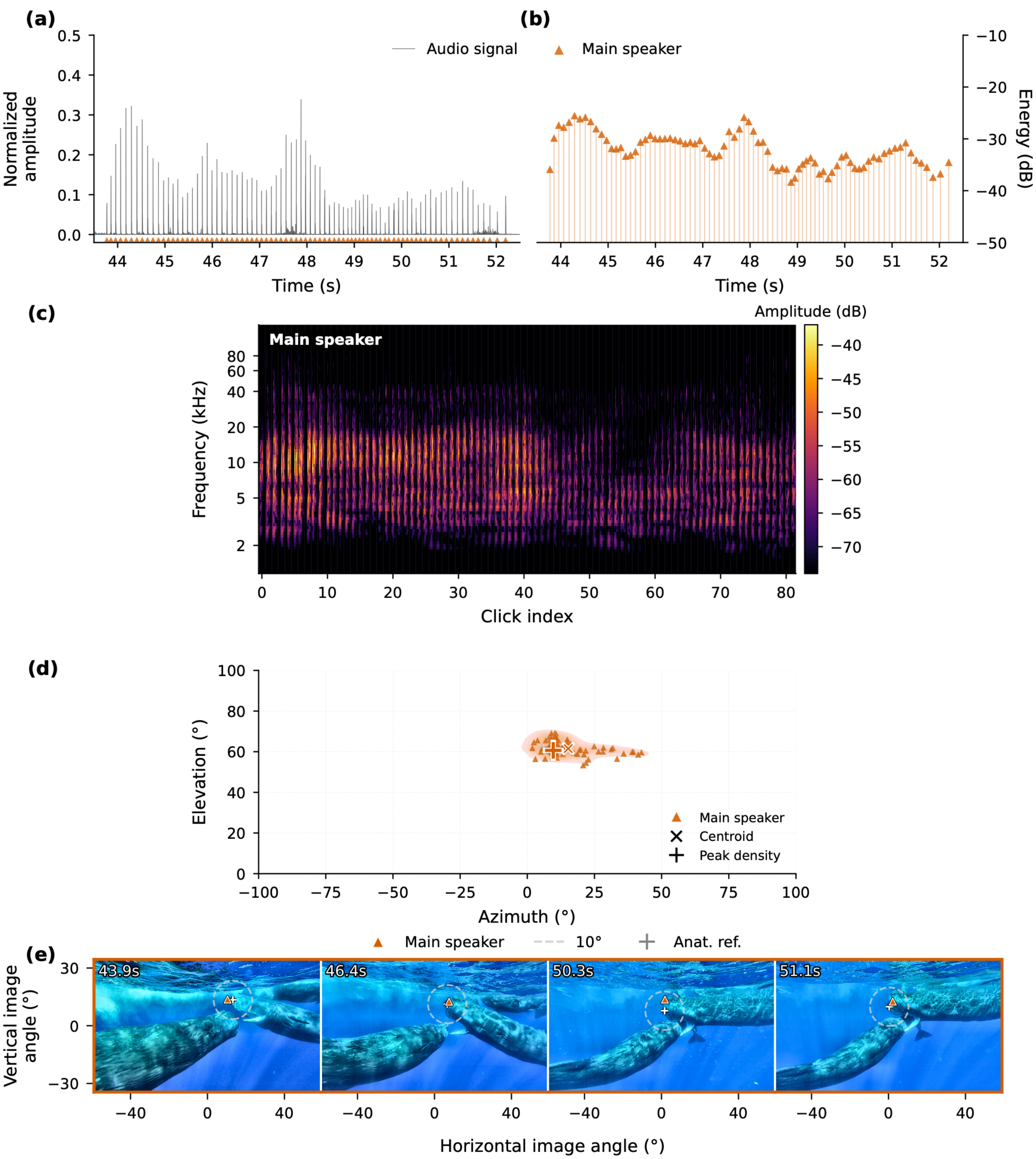

FIG. S4. (Color online). Multimodal source separation workflow across a moderately complex environment (Scene 7). Panels display: (a) normalized acoustic waveforms with marker overlays indicating click occurrence times; (b) temporal energy profiles (dB); (c) CQT spectral evolution computed from pre-whitened waveforms for single speaker; (d) acoustic spatial distribution (azimuth and elevation) via KDE, detailing centroids and peak densities; and (e) optical validation via projecting onto the synchronized video plane, displaying the 10° spatial tolerance threshold (dashed circles) around the anatomical reference (distal end of the junk). Orange triangles denote the single speaker (Ali, Track 7).

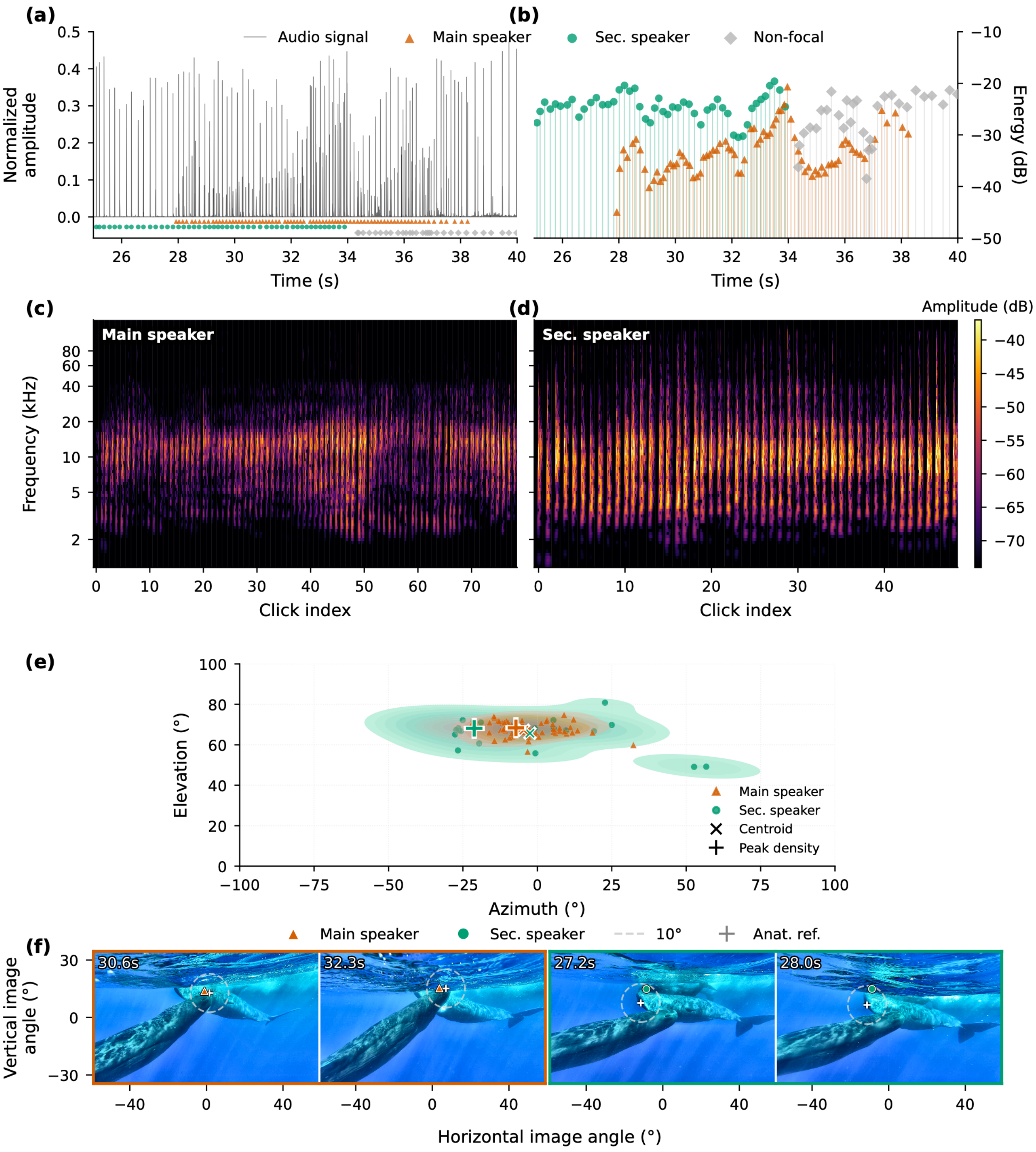

FIG. S5. (Color online). Multimodal source separation workflow across multi-emitter interaction (Scene 9). Panels display: (a) normalized acoustic waveforms with marker overlays indicating click occurrence times; (b) temporal energy profiles (dB); (c) and (d) CQT spectral evolution computed from pre-whitened waveforms for main and secondary speakers; (e) acoustic spatial distribution (azimuth and elevation) via KDE, detailing centroids and peak densities; and (f) optical validation via projecting onto the synchronized video plane, displaying the 10° spatial tolerance threshold (dashed circles) around the anatomical reference (distal end of the junk). Orange triangles denote the main speaker (Daren, Track 10); teal circles denote the secondary speaker (Ali, Track 11).

## ACKNOWLEDGMENTS

This work was supported by ULP-COCHLEA ANR-21-CE04-0020-01, ADSIL AID ANR-20-CHIA-0014, the Center of AI in Natural Acoustics http://cian.univ-tln.fr in the LIS Laboratory. We thank Longitude 181, Un Océan De Vie and Label Bleu Productions for fieldwork, video recordings and data support. We are especially grateful to René Heuzey, president of Un Océan De Vie, who initiated the missions, and to Axel Preud'homme and Navin Boodhonee for their valuable participation in the fieldwork. We also express our sincere gratitude for the support provided to the project "La Voix Des Cachalots" by the Mauritian Prime Minister's Office, Dr. Rezah Badal and his team, Chief Scientific Officer Mr. Satish Kadhun at the Albion Fisheries Research Centre (AFRC), Mr. Sachin Jootun and Miss Eliana Timol at the Mauritius Film Development Corporation (MFDC), and Director Miss Khoudijah Boodoo at the Tourism Authority. We are grateful to V. Gies, V. Barchasz and S. Marzetti (SMIoT Tech Platform Intelligent-Acoustics.com) for their help with the OPALE recorder. We also warmly thank D.S. Pace for support as a CSI member.

During the preparation of this manuscript, generative AI tools (Google Gemini and Anthropic Claude) were used as writing and formatting aids, specifically to assist with translation into English, language and style refinement, code development, debugging, and figure formatting. The scientific design of the methodology, its validation, and interpretation are the authors' work. All AI-assisted outputs were reviewed, tested, and validated by the authors, who assume the responsibility for the originality, accuracy, and scientific integrity of this publication.

## AUTHOR DECLARATIONS

### Conflict of Interest

The authors have no conflicts of interest to disclose.

### Ethics Approval

This study is based exclusively on non-invasive underwater observations of free-ranging sperm whales, involving no capture, tagging, or manipulation of the animals. The research adhered to the ASA Ethical Principles and followed the official charter for the responsible approach of marine mammals, ensuring that no avoidance behavior was induced. All marine scientific research was conducted in accordance with the Republic of Mauritius Maritime Zones (Conduct of Marine Scientific Research) Regulations 2017 (Government Notice No. 57 of 2017). Data collection for the 2023 campaign was authorized under a Marine Scientific Research Permit issued on 8 May 2023 by the Prime Minister's Office, Department for Continental Shelf, Maritime Zones Administration and Exploration (CSMZAE).

## DATA AVAILABILITY

The raw acoustic recordings analyzed in this study, together with the ground-truth datasets (manually validated acoustic click times and optical localizations for sessions 081356, 085315, and 085452), are openly available in the CIANSCAPE repository at https://sabiod.lis-lab.fr/pub/CIANSCAPE/JASA_Physeter_VocalSparring/ and in the Zenodo repository at https://doi.org/10.5281/zenodo.21891186. These repositories also contain the custom analysis notebooks (Python source code) for click detection, CQT extraction, spectro-temporal clustering, spatial localization, and video projection, documenting the methodology, as well as the rendered assignment videos (MP4) for all nine focal interaction scenes. The diarization metrics reported in the manuscript are computed from the provided validated reference files rather than regenerated by the notebooks. Compressed versions of two representative rendered scenes are included with the article as multimedia. The raw high-resolution video recordings are not deposited owing to their volume.

## APPENDIX

Table IV summarizes the click processing pipeline, from raw detection to the final focal subsets used for clustering and spatial localization (see Sec. II.C and II.E for methodological details). Figure 5 presents the multimodal source separation workflow for Scene 3, a representative case of a highly entangled multi-emitter interaction in which optical validation resolves the acoustic ambiguity.

TABLE IV. Complete click processing pipeline for the clustering (Panel A) and spatial localization (Panel B) stages, conducted in parallel across the three recording sessions. The two pipelines use distinct detection thresholds optimized for their respective objectives (Sec. II.C). Chronologically, all 1,122 preliminary clustered clicks from Panel A were initially submitted to spatial localization. The final subset of 637 focal clicks corresponds to the intersection of the 882 clustered clicks from the resolvable sessions (Scenes 1-4 and 7-9) with the 1,672 geometrically validated clicks from Panel B. Dashes in the final row indicate that session 085315 (Scenes 5 and 6; 240 clicks) was ultimately excluded from the final individual attribution due to unresolvable spatial ambiguities (Sec. III.C; Supplementary Table S3).

Panel A – Clustering pipeline:

| Stage | 081356 | 085315 | 085452 | Total |
|---|---|---|---|---|
| Automatic TKEO detections | 1,158 | 1,445 | 558 | 3,161 |
| Manually validated clicks | 850 | 1,295 | 510 | 2,655 |
| After adaptive envelope filter | 673 | 1,136 | 510 | 2,319 |
| Within 9 focal scenes (all emitters) | 545 | 295 | 402 | 1,242 |
| Clustered clicks – Scenes 5 & 6 (attribution unconfirmed) | - | 240 | - | 240 |
| Clustered focal clicks – retained sessions (excl. Sc. 5 & 6) | 537 | - | 345 | 882 |

Panel B – Localization pipeline:

| Stage | 081356 | 085315 | 085452 | Total |
|---|---|---|---|---|
| Strict threshold true positives | 665 | 1,201 | 442 | 2,308 |
| After loop closure + geometric residuals | 471 | 870 | 331 | 1,672 |
| Clustered clicks submitted to spatial projection (incl. Sc. 5 & 6) | 537 | 240 | 345 | 1,122 |
| Final focal clicks (intersection Panels A & B, excl. Sc 5 & 6) | 381 | - | 256 | 637 |

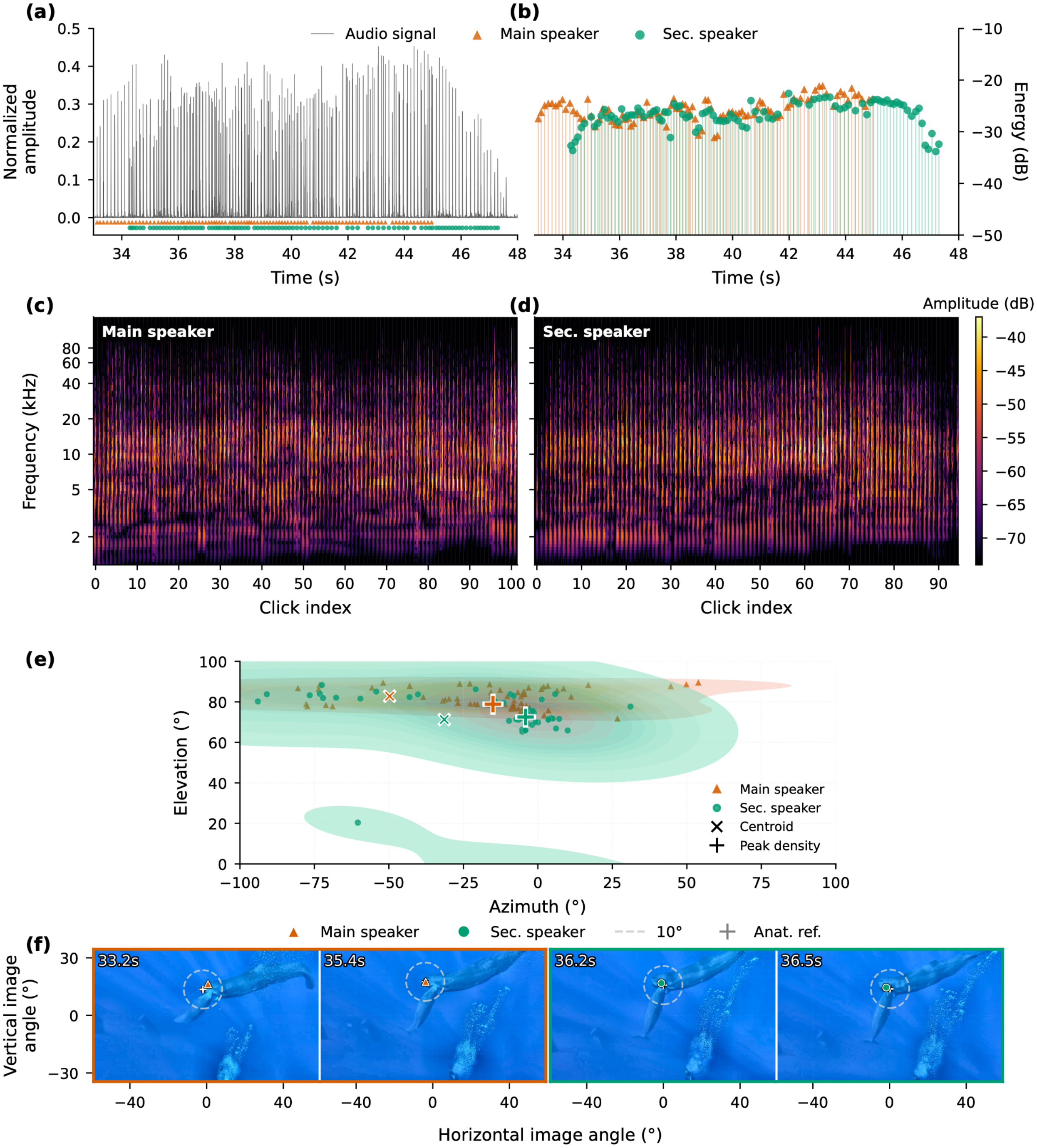

FIG. 5. (Color online). Multimodal source separation workflow across multi-emitter interaction (Scene 3). Panels display: (a) normalized acoustic waveforms with marker overlays indicating click occurrence times; (b) temporal energy profiles (dB); (c) and (d) Constant-Q Transform (CQT) spectral evolution computed from pre-whitened waveforms for main and secondary speakers; (e) acoustic spatial distribution (azimuth and elevation) via Kernel Density Estimation (KDE), detailing centroids and peak densities; and (f) optical validation via projecting onto the synchronized video plane, displaying the 10° spatial tolerance threshold (dashed circles) around the anatomical reference (distal end of the junk). Orange triangles denote the main speaker (Ali, Track 3); teal circles denote the secondary speaker (Daren, Track 4).